\documentclass[prl,twocolumn,
superscriptaddress,
 amsmath,amssymb,
 aps,
]{revtex4-2}

\usepackage{graphicx}
\usepackage{dcolumn}
\usepackage{bm}

\usepackage{hyperref}
\hypersetup{colorlinks=true,linkcolor=blue,urlcolor=blue,citecolor=blue}

\usepackage{ulem}

\usepackage[dvipsnames]{xcolor}

\newcommand{\gru}[1]{\textcolor{black}{#1}}
\newcommand{\vor}[1]{\textcolor{black}{#1}}

\begin{document}

\preprint{APS/123-QED}

\title{Collective Excitations and Stability of Nonequilibrium Polariton Supersolids}

\author{A. Grudinina}
\affiliation{Abrikosov Center for Theoretical Physics, Moscow Center for Advanced Studies, 141701 Moscow, Russia}
\affiliation{National Research Nuclear University MEPhI (Moscow Engineering Physics Institute), Kashirskoe shosse 31, 115409 Moscow, Russia}
\author{J. Cao}%
\affiliation{Abrikosov Center for Theoretical Physics, Moscow Center for Advanced Studies, 141701 Moscow, Russia}
\author{A. Kavokin}
\affiliation{Abrikosov Center for Theoretical Physics, Moscow Center for Advanced Studies, 141701 Moscow, Russia}
\affiliation{Russian Quantum Center, Skolkovo IC, Bolshoy boulevard 30 bld. 1, 121205 Moscow, Russia}
\author{N. Voronova}
\email{nsvoronova@mephi.ru}
\affiliation{National Research Nuclear University MEPhI (Moscow Engineering Physics Institute), Kashirskoe shosse 31, 115409 Moscow, Russia}
\affiliation{Russian Quantum Center, Skolkovo IC, Bolshoy boulevard 30 bld. 1, 121205 Moscow, Russia}
\author{A. Nalitov}
\email{anton.nalitov@gmail.com}
\affiliation{Abrikosov Center for Theoretical Physics, Moscow Center for Advanced Studies, 141701 Moscow, Russia}
\affiliation{Russian Quantum Center, Skolkovo IC, Bolshoy boulevard 30 bld. 1, 121205 Moscow, Russia}

\date{\today}

\begin{abstract}
Formation of nonequilibrium counterparts of supersolids, simultaneously characterized with spontaneous superfluid and crystalline order, was recently reported in incoherently pumped polariton condensates.
We investigate collective excitation spectra of this phase and explicitly demonstrate the emergence of gapless Nambu-Goldstone modes due to spontaneously broken continuous phase and translation symmetries. For the recent implementation of the polariton nonequilibrium supersolidity in semiconductor metasurfaces [D. Trypogeorgos {\it et al.}, \href{https://doi.org/10.1038/s41586-025-08616-9}{Nature {\bf 639}, 337 (2025)}], we demonstrate the key role of attractive polariton interactions, mediated by the excitonic reservoir, for stability of the supersolid phase. 
Performing a thorough numerical investigation, we identify the conditions for existence of the diagonal and off-diagonal long-range order in negative-mass nonequilibrium supersolids.

\end{abstract}

\maketitle
Supersolidity is characterized by the simultaneously broken gauge and translation continuous symmetries. While the gauge symmetry breaking results in the emergence of the off-diagonal long-range order and thus superfluidity, the breaking of translational invariance yields diagonal long-range order and the crystalline structure formation \cite{boninsegni2012supersolid,recati2023supersolidity}. 
Originally proposed in $^{4}$He~\cite{lifshitz1969supersolid, chester1970supersolidity, kirzhnits1971coherent, gross1957supersolid, legett1970supersolid}, supersolidity was then generalized to other superfluid systems displaying continuous translational symmetry breaking. Supersolid-like states were experimentally realized in Bose–Einstein
condensates (BECs) with spin–orbit coupling, where it takes the form of stripes~\cite{li2017supersolid}, atomic BECs inside optical resonators~\cite{leonard2017supersolid}, and in Bose gases with long-range dipolar interactions~\cite{tanzi2019supersolid,chomaz2019supersolid, bottcher2019supersolid}.

Recently, an alternative platform showcasing supersolidity was proposed~\cite{nigro2025supersolidity}, based on symmetry-protected optical bound-in-the-continuum (BiC) states which occur from two interfering resonances and enable light confinement~\cite{hsu2016BiC}. These infinite-lifetime photonic states can couple to matter excitations, such as excitons in transition-metal dichalcogenides~\cite{kravtsov2020BiC, zhang2018BiC}, quantum wells~\cite{bajoni2009BiC, ardizzone2022polariton}, or halide perovskites~\cite{masharin2025non, dang2020tailoring}, and form quasi-lossless hybrid light-matter quasiparticles called polaritons. Routinely studied in planar microcavities~\cite{QFL}, polaritons possess strong nonlinearity thanks to their exciton part and a small effective mass inherited from photons, and allow observation of coherent phenomena like Bose condensation~\cite{kasprzak_boseeinstein_2006}, superfluidity~\cite{amo2009superfluidity,lerario2017superfluidity}, or Josephson oscillations~\cite{BJJ_lagoudakis,BJJ_Abbarchi}, most of which are essentially non-equilibrium~\cite{bloch2022review}. Here, we turn to a new platform for exciton-polaritons in the quasi-BiC state: a patterned waveguide with embedded quantum wells, where ultralow-threshold Bose condensation of polaritons was recently demonstrated at the saddle point of the dispersion featuring a negative effective mass along the axis of the waveguide~\cite{ardizzone2022polariton,grudinina2023BIC}.  
For the very same system, formation of a polariton supersolid has been theoretically predicted~\cite{nigro2025supersolidity} and experimentally demonstrated~\cite{trypogeorgos2025emerging} exploiting parametric scattering (optical parametric oscillator effect, OPO~\gru{\cite{carusotto2005spontaneous,wouters2007OPO, claude2025observation}}) between the quasi-BiC condensate and adjacent modes that naturally arise in the 1D photonic crystal due to band folding. In the very recent \gru{works~\cite{meng2026supersolid, deng2026harnessing}, room-temperature supersolidity was obtained in a nanostructured waveguide and quantum confined microcavities utilizing the same OPO mechanism.} Furthermore, polaritons in a liquid-crystal cavity~\cite{muszynski2024observation, zhai2025electrically} demonstrate the formation of supersolid-like states analogous to the stripe phase in spin-orbit coupled BECs \cite{li2017supersolid, MartonePRL,Martone_rev} where the mechanism is based on the degeneracy of the ground state. 
Recently, polariton supersolidity was also reported in annular optically-induced traps~\cite{kozhevin2025supersolidity}, explained by the interplay of interactions between polaritons and incoherent excitons and the driven-dissipative nature of the system.

From the point of view of collective excitations, the formation of the supersolid state and breaking of the two continuous symmetries should be accompanied by the appearance of two gapless Nambu-Goldstone (NG) modes~\cite{PhysRevD.85.085010, watanabe2020counting}. 
Another signature of the supersolid transition is the formation of a band structure, along with the divergency of the structure factor at the Brillouin wave vector \cite{Martone_rev}. For patterned-waveguide polariton systems where the OPO-based supersolid features were reported~\cite{trypogeorgos2025emerging,meng2026supersolid}, although the instability analysis in the linear regime was provided~\cite{nigro2025supersolidity}, the demonstration of these hallmarks for the supersolidity formation is missing. 
In this Letter, we address the collective excitations of a polariton condensate in the quasi-BiC state above the second threshold related to the formation of the non-equilibrium supersolidity. We theoretically derive the excitation spectrum of the supersolid state and demonstrate the formation of two gapless NG modes which evidence the emergence of supersolidity. The presence of interactions between the coherent polariton subsystem and the incoherent reservoir is emphasized to be necessary for dynamical stability of the negative-mass polariton Bose condensate and the emergent supersolid phase. Our model offers a self-consistent interpretation of the observed dynamically stable light-matter phase and a framework for rigorously validating nonequilibrium supersolidity.

\begin{figure}[t!]
    \centering
\includegraphics[width=\linewidth]{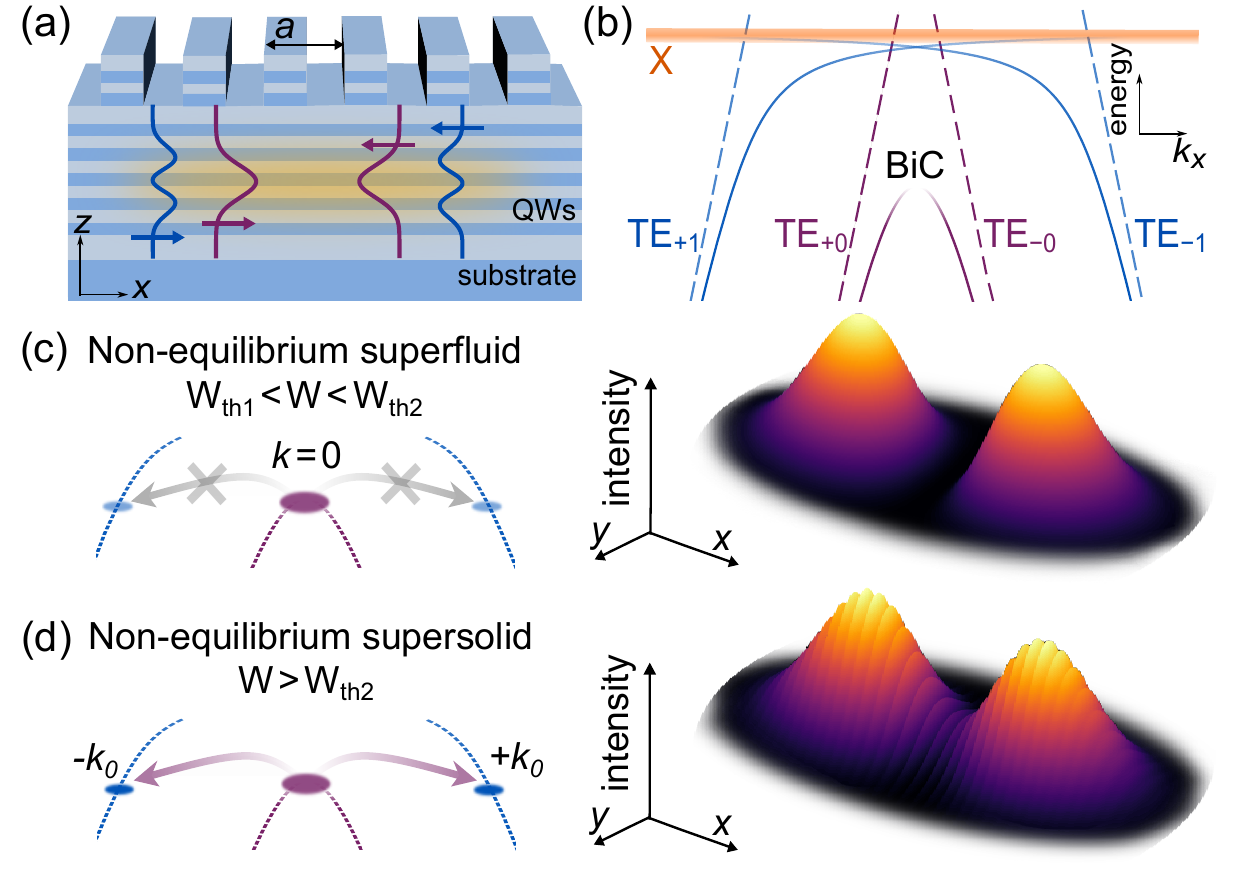}
    \caption{{\bf Schematic illustration of the system.} (a) The nanostructured waveguide with the grating period $a$. Layers of different colors show embedded QWs; the wavy lines illustrate fundamental photonic modes of the structure propagating in the $x$ direction, ${\rm TE}_{\pm 0}$ (purple) and ${\rm TE}_{\pm1}$ (blue). (b) The real parts of the dispersion laws $\varepsilon({\bf k})$ of the $0$ branch and $E_{\pm 1}({\bf k})$ of the $\pm1$ branches are shown with the solid purple and blue lines, respectively. The dashed lines correspond to the photonic modes sketched in (a). The thick orange line indicates the exciton resonance. (c) Sketch of the condensation at $k=0$ of the BiC branch (the dashed purple line) reached after the first threshold.
    (d) Sketch of the NESS formation above the second threshold: due to the OPO process, particles from the condensate in the $0$ branch (the purple cloud) scatter into the states at the $\pm1$ branches (the dashed blue lines) lying at the same energy with the condensate (the blue clouds). Coherent superposition of the modes with momenta $\pm {\bf k}_0$ results in a periodic modulation observed in the photoluminescence (PL) pattern. The 3D false-color images on the right in panels (c--d) show the PL intensity simulated using the Eqs.~\eqref{GPE} above the first (c) and second (d) thresholds.
    }
    \label{fig1}
\end{figure}

\paragraph{Theoretical model and mean-field analysis.} We consider a nanostructured optical waveguide with embedded quantum wells (QWs),  schematically shown in Fig.~\ref{fig1}(a). 
The consideration of ${\rm TE}_{\pm0}$ and ${\rm TE}_{\pm1}$ photonic modes coupled to QW excitons leads to a rich energy dispersion that consists of 8 branches (for details, see the Supplemental Material, SM~
\cite{footnote})
The model, however, can be reduced to just three lower modes shown in Fig.~\ref{fig1}(b).

The polariton Hamiltonian reads:
\begin{multline}\label{hamiltonian}
\hat{H} = \int d{\bf r}\biggl[ \hat{P}^{\dag}({\bf r}) \varepsilon(\hat{\bf k})\hat{P}({\bf r}) + \hat{P}^{\dag}_{+1}({\bf r}) E_{+1}(\hat{\bf k})\hat{P}_{+1}({\bf r}) \\+ \hat{P}^{\dag}_{-1}({\bf r}) E_{-1}(\hat{\bf k})\hat{P}_{-1}({\bf r})\biggr]
\!\!+ \!\frac{g}{2}\!\int \!d{\bf r} \hat{Q}^{\dag}({\bf r})\hat{Q}^{\dag}({\bf r})\hat{Q}({\bf r})\hat{Q}({\bf r}) \\+  \tilde{g} \int d{\bf r} \bigl[\hat{Q}^{\dag}({\bf r})\hat{Q}^{\dag}({\bf r})\hat{Q}_{+1}({\bf r})\hat{Q}_{-1}({\bf r}) + {\rm h.c.} \bigr],
\end{multline}
where $\hat{P}({\bf r},t)$ is the field operator describing lower lower polaritons occurring due to the coupling of excitons with ${\rm TE}_{\pm 0}$ photons, with the dispersion 
$\varepsilon({\bf k})$ 
(the main branch, denoted `0'), and $\hat{P}_{\pm1}({\bf r},t)$ are the field operators of lower polaritons arising from the coupling of excitons with ${\rm TE}_{\pm1}$  photons, with the dispersion 
$E_{\pm}({\bf k})$  (adjacent branches, `$\pm 1$'). 
The last two terms in~\eqref{hamiltonian} describe the interaction of polaritons via their exciton component. 
The exciton field operators corresponding to the $0$ and $\pm1$ modes are $\hat{Q}({\bf r},t) = \int d{\bf r}' X({\bf r}' -{\bf r}) \hat{P}({\bf r}', t)$ and $\hat{Q}_{\pm1}({\bf r},t) = \int d{\bf r}' X_{\pm1}({\bf r}' -{\bf r}) \hat{P}_{\pm1}({\bf r}', t)$, respectively, where $X({\bf r})$, $X_{\pm1}({\bf r})$ are the Fourier transforms of the exciton Hopfield coefficients $X_{\bf k}$, $X_{\pm1}({\bf k})$ whose shape is given in SM. 
\vor{The constant $g$ quantifies the interparticle interaction 
within the $0$ mode, which leads to the condensation in the BiC state at $k=0$ (see Fig.~\ref{fig1}(c); we refer to this phase as the non-equilibrium superfluid, NESF); $\tilde g$ describes interaction between the $0$ and $\pm1$ modes that results in the OPO process which is shown schematically in Fig.~\ref{fig1}(d).} 
We assume that the interaction between the polaritons on the $\pm1$ branches is negligible, since the population of these adjacent modes is much smaller than that of the main 0 mode~\cite{trypogeorgos2025emerging}. 

While the condensate at the BiC state is occupied thanks to the stimulated scattering of particles from the incoherent reservoir, macroscopic occupation of the adjacent ${\rm TE}_{\pm1}$ branches is due to the parametric scattering, resulting in the appearance of a coherent OPO state at a single energy $\mu$.
In the standard mean-field approach, the macroscopically occupied states can be treated separately.  
Averaging of the Heisenberg equations for the fields $\hat P$ and $\hat P_{\pm1}$ yields the Gross-Pitaevskii equations (GPEs) for the three components of the order parameter $\Psi_0 = \langle \hat{P}({\bf r}, t) \rangle $ and $ \Psi_{\pm} = \langle \hat{P}_{\pm 1}({\bf r}, t) \rangle$, which are derived in the SM. For a continuously driven system, we phenomenologically introduce the gain $+i W\Psi_0$ and gain saturation $-i \eta W |\Psi_0|^2$, which are associated with the adiabatically eliminated reservoir. In the above, $W$ is the effective pump power, $\eta=R/\gamma_R$, $R$ is the reservoir-to-condensate scattering rate and $\gamma_R$ is the reservoir decay rate (see SM for details).  We obtain:
\begin{subequations}\label{GPE}
    \begin{align}
    & i\hbar\frac{\partial}{\partial t}\Psi_{0}({\bf r}, t)=\varepsilon(\hat{\bf k})\Psi_{0}({\bf r}, t) + (g_{\rm eff}(W) - i \eta W)|\Psi_{0}|^2 \Psi_{0} \nonumber \\
    &\hspace{20pt} + g_R \frac{W}{R}\Psi_0 + 2 \tilde{g} \tilde{X} \Psi_{0}^* \Psi_{-}\Psi_{+} + i W\Psi_{0}, \label{GPE1}\\
    &i \hbar \frac{\partial}{\partial t}\Psi_{\pm}({\bf r}, t) = E_{\pm}(\hat{\bf k})\Psi_{\pm} + \tilde{g} \tilde{X}^{*} \Psi_{\mp}^*  \Psi_{0}^2. \label{GPE2}
\end{align}
\end{subequations}
Interactions between the main and adjacent modes are weighted with the coefficient $\tilde X$ whose shape is to be defined below. 
In Eq.~\eqref{GPE1}, we also take into account condensate-reservoir interactions $g_R n_R \Psi_0$, where $g_R$ is the interaction constant of polaritons with reservoir excitons and $n_R$ is the reservoir density. In the adiabatic reservoir limit, $n_R \approx W/R - (W/\gamma_{R})|\Psi_0|^2$. 
As a result, the effective interaction strength entering the GPE~\eqref{GPE1} reads $g_{\rm eff}(W) = g|X_0|^4 - g_R W/\gamma_R$, renormalized due to the condensate-reservoir interactions.

It is convenient to parameterize the order parameter components (for $0$ and $\pm1$ modes) as follows:
\begin{subequations}\label{parametrization}
    \begin{eqnarray}
   && \Psi_0({\bf r},t) =\sqrt{n_0} \vor{e^{i\phi_0}} e^{-i \mu t/\hbar},\\
    &&\Psi_{\pm}({\bf r},t) = \sqrt{n_{1}}e^{i \phi_{\pm}} e^{\mp i {\bf k}_0{\bf r}}e^{-i \mu t/\hbar},
\end{eqnarray} 
\end{subequations}
where $n_0$ \vor{and $\phi_0$ are} the $0$-mode density \vor{and phase}, 
$n_{1}$ and $\phi_{\pm}$ are the density and phases of the $\pm1$ modes, $\mu$ is the energy of the condensate. \vor{In a uniform mean-field configuration, without the loss of generality, $\phi_0$ can be taken as a reference and set to zero.}
The momenta at which the macroscopic occupation arises after the 2nd threshold are denoted as $\pm {\bf k}_0$ [see Fig.~\ref{fig1}(d)]. 
While generally the parametric scattering should include all the momenta on the same energy with the condensate, we note that the dispersion curvature along the $y$--axis is much smaller than that along $x$ 
(see SM). Hence we restrict ourselves to a quasi-1D model (along $k_x$ at $k_y=0$), and assume that the parametric scattering from the condensate involves exclusively the states with momenta $\pm {\bf k}_0 = \pm k_0 \hat{\bf x}$. 
The OPO process in Eqs.~\eqref{GPE} is therefore weighted with the following Hopfield coefficients: $\tilde{X} = |\tilde{X}|e^{i \delta \phi}= ({X}_{0}^*)^2 X_{+1}(-{\bf k}_0)X_{-1}({\bf k}_0)$. 
We emphasize that the periodicity that results from the appearance of the two modes with $\pm{\bf k}_0$ significantly differs from the periodicity of the 1D photonic crystal: $k_0$ is smaller than $\pi/a$ by almost an order of magnitude~\cite{trypogeorgos2025emerging}.

Importantly, for sufficiently large pump powers $W$, the renormalized interaction constant $g_{\rm eff}(W)$ can be negative-valued, which is crucial for the stability of the system. 
Since at relatively high pumping $g_{\rm eff}(W)$ switches its sign, in combination with the negative effective mass along the $k_x$--direction, interaction recovers its repulsive nature, stabilizing the condensate and preventing it from the collapse into droplets in real space~\gru{\cite{PhysRevB.89.235310,PhysRevB.91.085413, estrecho2018single, baboux2018unstable}}. Similar effect of the negative effective mass was obtained in Refs.~\cite{gianfrate2024reconfigurable,Sigurdsson2024} where it was shown to result in the effective pump-induced repulsive potential acting as a trap. It should be noted that \gru{condensed and supersolid phases possess instabilities of different nature, but the change of the interaction sign leads to the stabilization of both (for details, see End Matter, EM)}. Without the condensate-reservoir interactions, the model would demonstrate inconsistencies with the experiment, producing unstable excitation spectra (see SM).

The expression for the condensate energy $\mu$ versus pump power reads $\mu(W)= g_{\rm eff}(W)n_0 + g_R W/R$. 
In this case the scattering momenta $\pm {\bf k}_0$, which are defined from the equation ${\rm Re}E_{\pm}(\mp{\bf k}_0) = {\mu}$, start to depend on pump power: ${\bf k}_0={\bf k}_0(W)$. 
At the same time, the condensate blueshift of order $\sim 1$~meV results in less than 2\% change of the absolute value of $k_0$. Hence it is reasonable to assume that in the range of $k_0$ swept by the considered change of $W$, the losses of the $\pm1$ modes are weakly dependent on $k$ and can be approximated as constant ${\rm Im}E_{\pm}({\bf k}) \approx {\rm const}$. 
\begin{figure}[t!]
    \centering
\includegraphics[width=1.\linewidth]{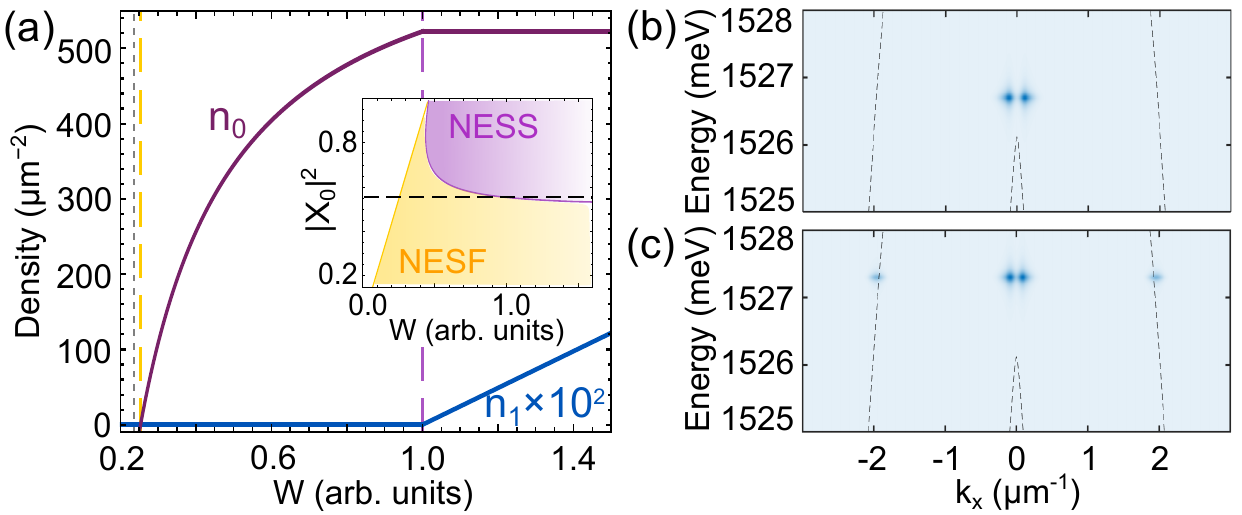}
    \caption{(a) Density of the order parameter components $n_0$ (purple) and $n_1$ (blue) vs. the pump power $W$. The dotted gray line indicates the value of $W$ at which the effective interaction changes sign, $g_{\rm eff}(W)<0$, and the interaction of negative-mass polaritons effectively becomes repulsive; the yellow and lilac dashed lines mark the first and second thresholds. The inset shows phase boundaries for the condensed phase (NESF, yellow) and the supersolid phase (NESS, purple). The black dashed line indicates the value of the 0-mode Hopfield coefficient at which the calculations presented in the paper are performed. The PL intensity in reciprocal space for the NESF (b) and the NESS (c) phases calculated at $W = 1.2~W_{\rm th1}$ and $W =1.2~W_{\rm th2}$, respectively. 
    Parameters: $g = \tilde g = 2.5$~$\mu$eV$~\mu$m$^{2}$, $\eta = 1.4\times10^{-3}$~$\mu$m$^{2}$, $R=6.7\times 10^{-3}$~$\mu$eV$~\mu$m$^{2}$, $g_R = g |X_0|^2$. Other parameters of the model are given in the SM. }
    \label{fig2}
\end{figure}

The GPEs~(\ref{GPE}) allow to derive the equations for the densities $n_0$, $n_{1}$, and phases $\phi_{\pm}$ (see SM). The proposed model reproduces the two-threshold behavior: the first threshold corresponds to the macroscopic population of the main 0--mode, while the second threshold indicates that the adjacent modes start to be populated. Above the second threshold, with the onset of nonzero population of the adjacent modes, the stable configurations read:
\begin{subequations}\label{densities}
    \begin{align}
    &n_0 = \frac{|{\rm Im} E_{\pm}(\mp{\bf k_0})| }{\tilde{g}|\tilde{X}|}, \quad n_1 = \frac{ W-\gamma- \eta W n_0}{2 \tilde{g}|\tilde{X}|} ,\\
   &\phi_{+} + \phi_{-} + \delta \phi = -\frac{\pi}{2}, \label{phases}
\end{align}
\end{subequations}
where $\gamma = |{\rm Im}\varepsilon(0)|$, $\delta\phi$ is the phase of $\tilde X$ (see above). Importantly, the fixed relation~\eqref{phases} between the phases $\phi_{+}$ and $\phi_{-}$ results in the OPO process working as loss for the main mode and gain for the adjacent modes.

In Fig.~\ref{fig2}(a) we plot the dependence of the main and adjacent modes densities $n_0$ and $n_1$ on pump power, according to Eqs.~(\ref{densities}). The inset of Fig.~\ref{fig2}(a) shows the phase diagram in terms of the 0-mode Hopfield coefficient $|X_0|^2$ and pump power $W$, illustrating the BEC-supersolid phase transition---we refer to these phases as the non-equilibrium superfluid (NESF) and non-equilibrim supersolid (NESS) states.
One sees that in the case of photonic detunings, $|X_0|^2<0.5$, the second threshold is not observed: since the OPO nonlinearity is weighted with the exciton fraction, for photon-like polaritons the scattering into adjacent modes is weak. The threshold values are obtained in the SM:
\begin{equation}
W_{\rm th1} = \gamma, \quad
W_{\rm th2} = \gamma\tilde{g}|\tilde{X}|/[\tilde{g}|\tilde{X}| - \eta |{\rm Im}E_{\pm}(\mp {\bf k}_0)|].  \label{th2}
\end{equation} 
Fig\gru{s}.~\ref{fig2}(b--c) show the photoluminescence (PL) profile in $k$--space obtained from numerical simulations of Eqs.~\eqref{GPE}, for the cases $W_{\rm th1}<W<W_{\rm th2}$ (b) and $W>W_{\rm th2}$ (c).

\paragraph{Excitation spectrum.} We define the fluctuations above the stable configurations of the order parameter components in the following way:
\begin{subequations}\label{par1}
    \begin{align}
    &\Psi_{0}({\bf r},t) = e^{-i \mu t/\hbar}(\sqrt{n_0} + \delta\Psi_0({\bf r},t)),\\
   &\Psi_{+}({\bf r},t) = e^{i \phi_{+}}e^{-i \mu t/\hbar} (\sqrt{n_1} e^{-i {\bf k}_0 {\bf r}} + \delta\Psi_{+}({\bf r},t)), \\
   &\Psi_{-}({\bf r},t) = e^{i \phi_{-}} e^{-i \mu t/\hbar} (\sqrt{n_1} e^{i {\bf k}_0 {\bf r}}+ \delta\Psi_{-}({\bf r},t)).
\end{align}
\end{subequations}
Since the order parameter is spatially modulated, using the Bloch theorem the fluctuations can be expressed as:
\begin{eqnarray}
   &\delta\Psi_{j}({\bf r},t)  = \gru{\sum\limits_{{\bf k}, l} e^{i {\bf k}{\bf r} - i \omega_{l,k}t} \sum\limits_{m\in \mathbb{Z}}u^{(j)}_{l, k+ m k_1} e^{i m k_1 x}} \nonumber \\ & \hspace{40pt}\gru{- \sum\limits_{{\bf k}, l} e^{-i {\bf k}{\bf r} + i \omega^{*}_{l,k}t}\sum\limits_{m \in \mathbb{Z}}v^{(j)*}_{l, k+ m k_1} e^{-i m k_1 x}}, 
\end{eqnarray}
where $\omega_{l,k}$ denote the dispersions of elementary excitations,  $j = \{0, +1, -1\}$, $u^{(0)}_{l,k}, v^{(0)}_{l,k}$ and $u_{l,k}^{(\pm)}, v_{l,k}^{(\pm)}$ are the Bogoliubov amplitudes corresponding to the 0 and $\pm1$ modes, respectively, $l$ is the band number, and $k_1$ is the wavevector characterizing the periodicity of modulations. In our case, $k_1=k_0$. 
As a result, the excitation spectra demonstrate the band structure formation.
This parameterization leads to the fluctuation matrix of infinite dimension. 
In the long-wavelength limit, this matrix can be truncated to account for several first bands (see SM for details).

\begin{figure}[t]
    \centering
\includegraphics[width=1\linewidth]{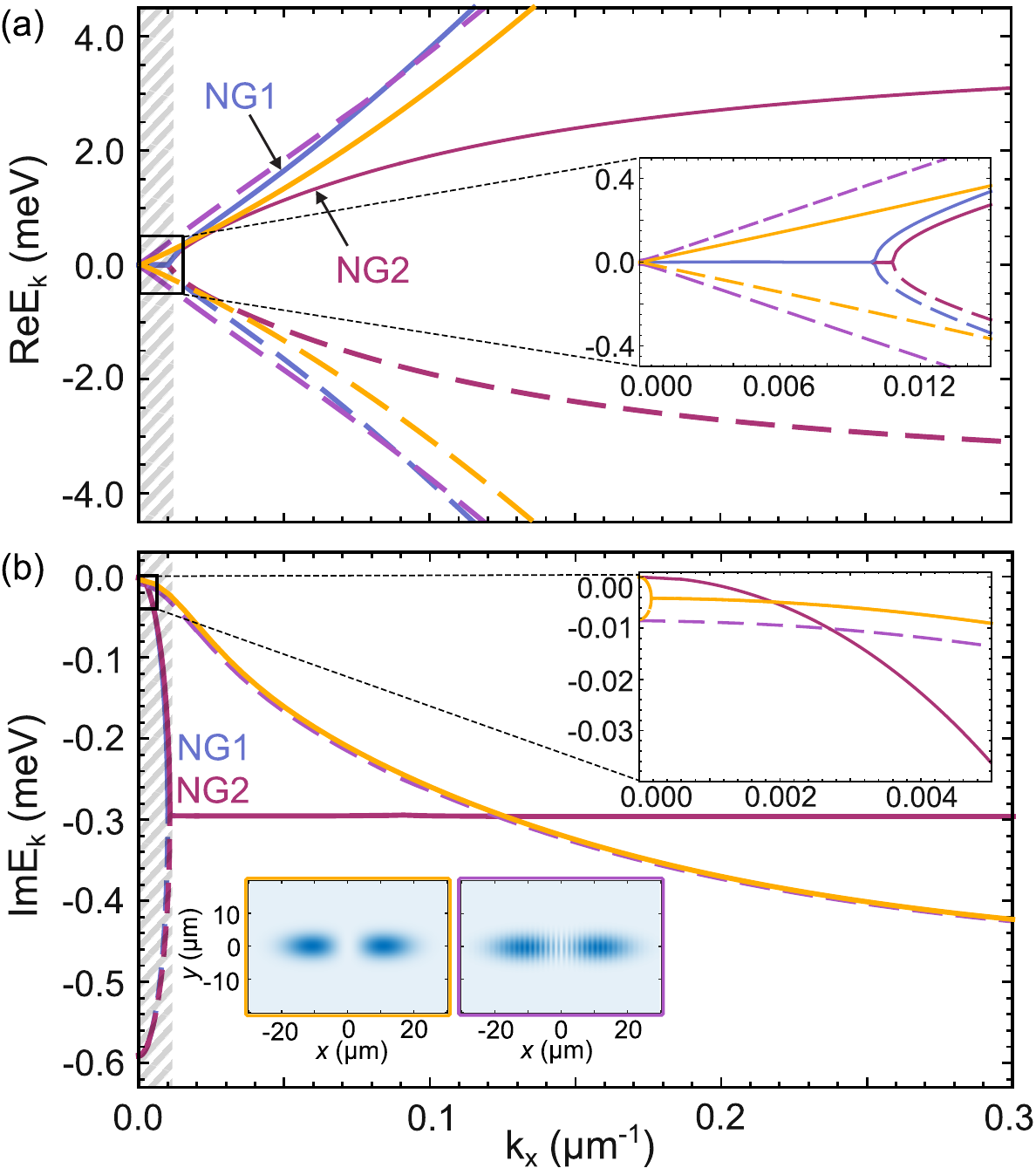}
    \caption{{\bf The excitation spectra $E_{\bf k}$ as a function of $k_x$}. 
    Real (a) and imaginary (b) parts of $E_{\bf k}$ in the first band for the NESS (the purple, magenta, and lavender lines)  and NESF (the yellow lines) phases. The solid lines in panels (a, b)  correspond to the NG modes (see text), the dashed lines indicate the gapped modes.  
    Parameters above the 2nd threshold: $W = 1.05~W_{{\rm th}2}$, $\mu = 1$~meV, $n_0=520~\mu$m$^{2}$, $n_1/n_0=2\times 10^{-4}$; below the 2nd threshold: $W = 2.5~W_{{\rm th}1} = 0.64~W_{{\rm th}2}$, $n_0 = 420~\mu$m$^{-2}$, $\mu=0.9$~meV. The insets in panels (a) and (b) show the smaller momentum range enlarged. The color-framed insets in (b) show the simulated PL distributions in real space before (yellow) and after (purple) the second threshold. Hatched areas in panels (a) and (b) show the characteristic momenta corresponding to the diffusive parts of the excitation spectra. Other parameters as in Fig.~\ref{fig2}.
    }
    \label{fig3}
\end{figure}

The first band of excitation spectra in the vicinity of $k=0$ is shown in Fig.~\ref{fig3}: the purple, lavender, and magenta lines display the Bogoliubov dispersions after the second threshold, when the modulated density pattern is formed for the stationary configurations~(\ref{densities}). One sees six branches: in $k\to0$ limit, ({\it i}) two branches are gapless with the negative imaginary parts which tend to zero (the  magenta and lavender solid lines), ({\it ii}) two branches have zero real parts and non-zero imaginary parts (the magenta and lavender dashed lines), and ({\it iii}) the remaining pair is gapped, with the finite negative imaginary part (the dashed purple lines). 
At large $k$, on the other hand, the shape of all six branches is governed by the shape of the single-particle dispersions. Importantly, the two gapless modes can be interpreted as 
Nambu-Goldstone modes \gru{(${\rm Re}E_{\bf k}$, ${\rm Im}E_{\bf k}\to0$ \vor{at ${\bf k}\to0$}, see EM for details)}. The narrow diffusive parts of NG modes in the vicinity of $k=0$ (see the hatched area in Fig.~\ref{fig3} and the insets for a larger view) \gru{have the scale of $\Delta k = 0.01~\mu$m$^{-1}$}; 
for realistic condensates of the sizes $<1$~mm, this range of wavevectors is not accessible. 

To compare, the same analysis for $W<W_{\rm th2}$, when $n_1=0$ and $n_0 = (W -\gamma)/(\eta W)$, reveals one gapless and one gapped mode (the yellow solid and dashed lines in Fig.~\ref{fig3}, respectively). The sound-like mode is also diffusive only in a very narrow range of near-zero momenta, indicating the breakdown of long-range order, typical for non-equilibrium condensates.
We see that when crossing the second threshold, the excitation spectrum changes from consisting of two branches to that with six branches, and the simultaneous appearance of the two gapless NG modes instead of one can be interpreted as evidence of the supersolid state formation. Note that we have considered a scalar problem (i.e., neglected the polariton pseudospin degree of freedom). For the spinor case, the excitation spectra are expected to be richer~\cite{PhysRevLett.97.066402}.

\gru{The long-wavelength analysis \vor{presented in EM} shows that the two gapless modes correspond to the fluctuations of $\phi_0$ and $\phi_{+} - \phi_{-}$.} 
\vor{This is as expected} from the symmetry point of view: the Hamiltonian~\eqref{hamiltonian} is invariant under the 
transformation $\Psi_{0} \to \Psi_{0} e^{i \beta}$,  $\Psi_{\pm} \to \Psi_{\pm} e^{i \beta}$ (so that one of the observed NG modes can be attributed to the {\it gauge} 
symmetry breaking); at the same time, the phases of the adjacent modes $\phi_{\pm}$ relative to the phase \vor{$\phi_0$} of \vor{the main mode} 
are only bound by the relation~\eqref{phases}. 
A particular choice of $\phi_{\pm}$ leads to the displacement of the sinusoidal modulations in the density profile. Therefore the second gapless mode can be attributed to the spontaneous breaking of $U(1)$ symmetry with respect to $\phi_{\pm}$, and this random choice of the relative phase results in the {\it translational} symmetry breaking~\cite{wouters2007OPO}.

\paragraph{Conclusions.} We examined the collective excitation spectrum of the exciton-polariton supersolid phase that was recently experimentally reported~\cite{trypogeorgos2025emerging}. 
We have shown, in agreement with the experimental evidence, that despite the negative effective mass, the BiC-based polariton system can support stable phases with broken gauge and translational symmetries. 
The scenario of the nonequilibrium supersolid (NESS) formation is shown to be different from that in atomic BECs. Namely, the presence of the incoherent exciton reservoir modifying the GPE mean-field approach is required, and certain conditions on the system parameters should be satisfied in order to guarantee the stability of elementary excitations. 

Importantly, the role of incoherent reservoir is twofold. First, the resulting effective pumping and gain saturation in the model with the OPO scattering mechanism leads to the two-threshold behaviour. Second, interaction of the reservoir with the condensate particles, apart from producing additional blueshift, can also affect substantially the polariton interaction constant that starts to be pump-power dependent and changes sign at a certain power. Note that the renormalization of the polariton interaction constant due to reservoir has been previously discussed in literature~\cite{PhysRevB.89.235310,PhysRevB.91.085413} and that the switching from polariton repulsion to attraction has been experimentally achieved~\cite{estrecho2018single, baboux2018unstable}. Here, in combination with the negative effective mass, the sign-reversal of interaction is crucial for 
\vor{stability of the system against perturbations, both in the condensed phase and in the supersolid state, even while the stabilization mechanisms are different.}

As a smoking gun of the emergence of NESS, our theory predicts that above the second threshold, the band structure is formed in the excitation spectrum and two gapless Nambu-Goldstone modes appear. By contrast, below the second threshold, only one NG branch arises according to spontaneously broken $U(1)$ phase symmetry. 
Our model for computing collective excitation spectra provides a general framework for benchmarking a wide class of nonequilibrium supersolids.

\begin{acknowledgments}

The work is funded by the Russian Science Foundation grant Project No.
25-22-00886 (\url{https://rscf.ru/en/project/25-22-00886/}).
\end{acknowledgments}


\subsection{End Matter}

\vspace{-18pt}
\vor{\paragraph{Long-wavelength limit.}
To gain insight into the nature and stability of the arising gapless modes, we look more closely at the phase and density 
perturbations on top of the stationary state~\eqref{parametrization}.
Truncating the Bogoliubov fluctuation matrix to $6\times6$ and rewriting it in terms of low energy $(k\to 0, \,\omega \to 0)$ fluctuations (see details in SM Sec.~III), one obtains 
\begin{equation}\label{eq_phases}
    \begin{split}
        \hbar \omega \!\left(2-\frac{n_0}{n_1}\right)\delta \phi_0 & = \hbar v k \, (\delta \phi_+  -\delta \phi_-), \\
        \hbar \omega (\delta \phi_+  -\delta \phi_-) & = 2\hbar v k \, \delta \phi_0,
    \end{split}
\end{equation}
where the stationary values of $n_0$ and $n_1$ are given by Eqs.~\eqref{densities}, and $\pm v$ is the linear slope coefficient of the $\pm1$ modes at $k_x \sim \mp k_0$. 
Higher corrections in $k$ and $\omega$ to Eqs.~\eqref{eq_phases} are provided in the SM.
%
One sees that at $W>W_{\rm th2}$ (when $n_1\neq0$), the $k\to0$ regime of the system is described by the coupled equations~\eqref{eq_phases} for the BiC-mode phase $\phi_0$ and the relative phase $\phi_+ - \phi_-$ of the $\pm 1$ modes, which yield the two gapless branches of the dispersion
\vspace{-5pt}
\begin{equation}\label{omega_k}
    E_{{\bf k}\to0}(W>W_{\rm th2}) = \pm i \hbar v\sqrt{\frac{2n_1(W)}{n_0-2n_1(W)}}\,k.
\vspace{-5pt}
\end{equation}
Hence, for the case considered here ($n_1\ll n_0$), in terms of the phase and amplitude fluctuations, the two gapless modes 
in the limit $k\to0$ correspond to phase fluctuations $\delta\phi_0$ and 
$\delta \phi_+ -\delta\phi_-$, and 
can be addressed as Nambu-Goldstone modes. At the same time, one sees that at finite $k$, one of these NG modes demonstrates non-zero positive imaginary part (not seen in Fig.~\ref{fig3}) and should, evidently, become unstable---\textit{in contradiction} with the experimental observations.}
%

\vor{Importantly, the obtained $k\to0$ dispersions~\eqref{omega_k} above the second threshold, while corresponding to fluctuations of the phases, are {\it independent} of the effective mass, and hence the origin of this intrinsic instability is {\it not} connected to the negative sign of $m$. This is in stark contrast with the case $W<W_{\rm th2}$ ($n_1\equiv 0$) and/or the absence of OPO ($\tilde g$ = 0), where 
the long-wavelength limit yields
the well-know diffusive low-$k$ dispersion~\cite{QFL}
\begin{equation}\label{Ek_Bog}
    E_{{\bf k}\to0}(W< W_{\rm th2}\text{ or }\tilde g=0) = -i \,\frac{\hbar^2 g_{\rm eff}(W)}{2\eta W m}k^2
\end{equation}
which, indeed, displays an instability (positive imaginary part) 
when the mass and interaction constant have different signs.
We conclude therefore that the long-wavelength instability above the second threshold 
has a different origin and it emerges due to the OPO process itself (regardless of the `sign' instability occurring due to ${\rm sgn} \, g_{\rm eff}\neq{\rm sgn} \, m$).} 

\vor{\paragraph{Stabilization in the low-$k$ region.} As we have emphasized above, the effective interaction constant $g_{\rm eff}(W)$ 
may be negative-valued, which leads effectively to the stabilization of the condensate (NESF phase), despite the negative effective mass along the $k_x$ direction. This can be immediately seen from the  spectrum in Eq.~\eqref{Ek_Bog}.} 
\vor{More subtle is the effect of the negative-valued $g_{\rm eff}$ in combination with the negative mass on the elementary excitation spectra above the second threshold (in the NESS phase), 
which have the long-wavelength asymptotics~\eqref{omega_k}. The branch with ${\rm Im}E_{{\bf k}\to0}>0$, in general, indicates that the chosen uniform background state is dynamically unstable to long-wavelength perturbations. 
The origin of this instability is that the parametric gain of the OPO process overpowers local dissipation; however, it is important to think about the values of the wavevectors corresponding to the growing modes. If stationary ($\omega=0$) oscillating solutions would exist for small non-zero $k$, this would indicate that the wavelength of such ``dangerous'' perturbations is limited from below, and that the exponential growth of fluctuations in time must saturate due to the system's nonlinearities, bringing the system to a spatially modulated steady state 
(the so-called stationary patterned state~\cite{pattern_book}).
}

\vor{To follow the behavior of phase and density fluctuations at small, but non-zero momenta we plug the stationary configuration~\eqref{parametrization} into 
Eqs.~\eqref{GPE} 
and approximate ${\rm Re}\,\varepsilon(\hat{\bf k})  \simeq -\hbar^2\partial_{xx}^2/2m$, ${\rm Re}E_\pm(\hat{\bf k})  \simeq {\rm Re} E_{\pm}(\mp {\bf k}_0) \mp i \hbar v \partial_x$, 
where $m$ is the (negative) effective mass of the BiC mode in the $k_x$-direction (see Sec. IIA of SM). 
Changing variables from $\phi_\pm$ to the relative phase of the adjacent modes $\Phi = \phi_+ \!-\phi_-$ and 
the irrelevant total phase $\zeta = \phi_+ \!+\phi_- \!\! - \! 2\phi_0$, we shift the mean-field values 
$n_0\to n_0 + \delta n_0$, $n_\pm \to n_1 + \delta n_{\pm}$, $\phi_0 \to 0 +\delta \phi_0$, $\phi_\pm\to\phi_\pm +\delta \phi_{\pm}$. Neglecting spatial derivatives of $\delta n_0$ in the long-wavelength limit, one gets for the fluctuations
\begin{equation} \label{eq_fluct} 
    \begin{split}
    \hbar\partial_t \delta n_0 & \! = \!- n_0\Bigl[\tfrac{\hbar^2}{m}\partial_{xx}^2\delta\phi_0 \!-\! 2\eta W\delta n_0 \!-\! 2\tilde g|\tilde X|\bigl(\delta n_+ \!\!+\! \delta n_-\!\bigr)\Bigr],\\
    \hbar\partial_t \delta n_\pm  & \! = \! 2\tilde g|\tilde X|n_1 \delta n_0 \!-\! \bigl(\pm\hbar v\partial_x \!\!+\! \tilde g|\tilde X|n_0\bigr)\delta n_\pm \!\!+\! \tilde g|\tilde X|n_0 \delta n_\mp,\\
    \hbar\partial_t\delta\zeta & = -\hbar v\partial_x\delta \Phi \!+\! 2g_{\rm eff}(W)\delta n_0 \!-\! 2 \tilde g|\tilde X|(n_0 \!-\! 2n_1)\delta\zeta,\\
    \hbar\partial_t\delta\phi_0 & = - g_{\rm eff}(W)\delta n_0 - 2\tilde g|\tilde X| n_1\delta\zeta, \\
    \hbar\partial_t\delta \Phi & = -\hbar v \partial_x \delta\zeta - 2\hbar v\partial_x\delta\phi_0.
    \end{split}
\vspace{-6pt}
\end{equation}
Note that it is crucial to retain the quantum-pressure term $\partial_{xx}^2\delta\phi_0$ (``restoring force'') and the gradients of the adjacent-mode densities $\partial_x\delta n_\pm$ (``group velocity''). 
In the adiabatic approximation for the densities and the total phase, this allows to establish the connection 
$$
\partial_{xx}^2\delta\phi_0 = \frac{g_{\rm eff}}{4\tilde g|\tilde X|n_1} \partial_{xx}^2\delta n_0.
\vspace{-2pt}
$$
Then, combining the equations for the density fluctuations, one finally arrives at the fourth-order differential equation for $\delta n_0$:
\begin{equation}
C_1\delta n_0^{\prime\prime\prime\prime} + C_2\delta n_0^{\prime\prime}     - C_3\delta n_0 =0 
\end{equation}
with 
\vspace{-5pt}
$$C_1 = \frac{\hbar^2}{8mn_1}\frac{g_{\rm eff}}{\tilde g^2 |\tilde X|^2}, \,\, C_2 = \frac{\eta W}{\tilde g|\tilde X|}, \,\, C_3 = \frac{8 \tilde g^2 |\tilde X|^2 n_1n_0}{\hbar^2 v^2}.
\vspace{-2pt}
$$
Substituting $\delta n_0\propto e^{ikx}$, one finds immediately
\begin{equation}\label{kc}
    k^2 = \frac{C_2}{2C_1}\pm\sqrt{\frac{C_2^2}{4C_1^2} + \frac{C_3}{C_1}}.
\end{equation}
One sees that, since $C_{2,3}$ are strictly positive, $C_1>0$ results in the existence of real $k$ and harmonic solutions corresponding to a stable, periodic density wave in the bulk of the system (the `plus' root), while the `minus' root yields real exponential solutions that fulfill the boundary conditions (acting as evanescent waves pinned to the edges of the system). On the contrary, if $C_1<0$, only the unbounded solutions ($k^2<0$) exist, and no stable pattern formation is supported.}

\begin{figure}[h!]
\includegraphics[width=\linewidth]{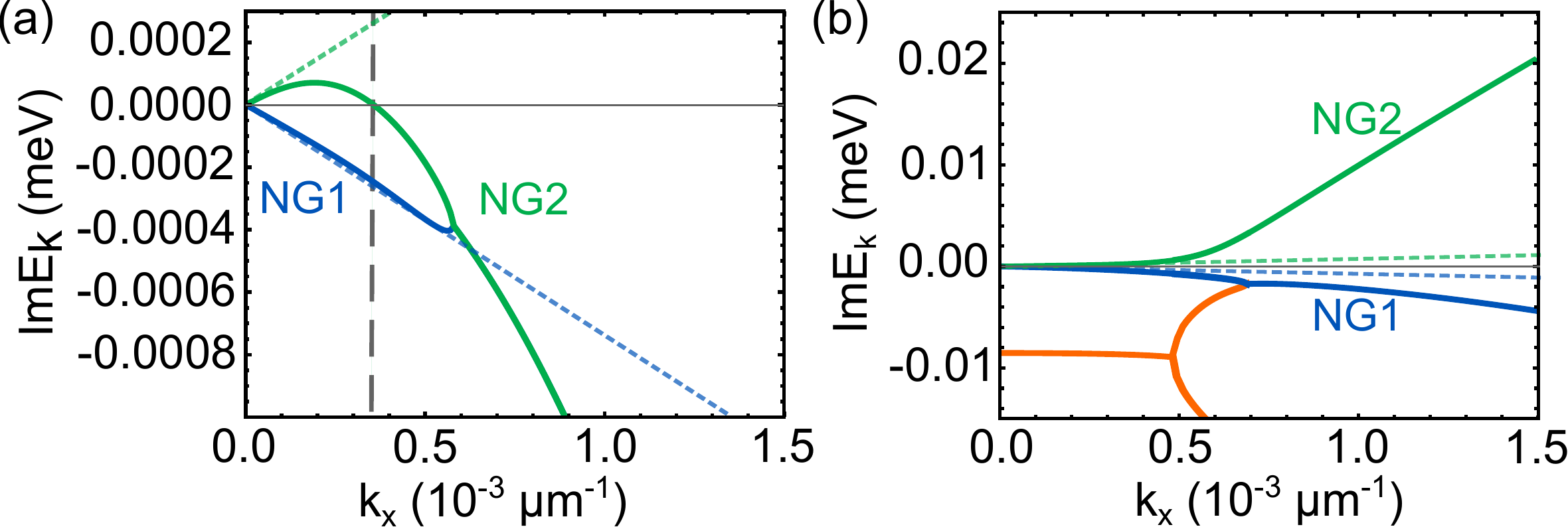}
    \caption{\vor{The low-$k$ region of the excitation spectra ${\rm Im}E_{\bf k}$ above the second threshold. 
    (a) Zoom-in view of the inset of Fig.~\ref{fig3}(b) with $g_{\rm eff}<0$. The OPO instability becomes confined to the region $k<k_c$ (the vertical dashed line). (b) Same for the ``usual'' case $g_R=0$ and $g_{\rm eff} = g|X_0|^4>0$. The dotted lines in both panels show the analytic result of Eq.~\eqref{omega_k}, confirming linearisation at $k\to0$.}}
    \label{SFig_new}
\end{figure}

\vor{By definition, $\text{sgn}[C_1] = \text{sgn}[g_{\rm eff}/m]$, and therefore for the negative-mass case that we consider, the renormalization of interaction constant by the reservoir interactions, $g|X_0|^4 \to g_{\rm eff}$, and the sign reversal of $g_{\rm eff}$ is of crucial importance for the system's stability above the second threshold. In Fig.~\ref{SFig_new}, we plot the comparison of the imaginary part of the  spectra in the small-$k$ region, for the cases when interactions with the reservoir cause the effective change of the interaction sign [$g_R\neq 0$, $g_{\rm eff}<0$, Fig.~\ref{SFig_new}(a)] and when the reservoir interaction is absent [$g_R=0$, Fig.~\ref{SFig_new}(b); the full view of the spectra in this case are provided in SM]. 
The harmonic spatial solutions that---in the negative-mass case---exist only for $g_{\rm eff}<0$, correspond to the critical wavevector $k_c$ that forms the boundary between the dynamically unstable long-wavelength regime and the regime of stability ${\rm Im}E_{\bf k}\leq0$ (see the dashed line in Fig.~\ref{SFig_new}(b)). For $k>k_c$, the terms $\sim k^2,k^4$ arising from the quantum pressure provide a massive hydrodynamic restoring force that overwhelms the parametric gain. Estimating $k_c$ from Eq.~\eqref{kc} with the parameters used in this work, we get $k_c\lesssim 10^{-3}\mu$m$^{-1}$ even for large pump powers $W\sim5W_{\rm th2}$, while for the power used in Fig.~3 
($1.05 W_{\rm th2}$) its value $k_c\sim10^{-4}\mu$m$^{-1}$ corresponds to the spatial scale of tens of millimeters, 
and 
cannot produce an instability in current finite-spot experiments~\cite{pethick_book, stoof_book}.
}

\end{document}


\title{Collective Excitations and Stability of Nonequilibrium Polariton Supersolids\\
Supplemental Material}

\author{A. Grudinina}
\affiliation{Abrikosov Center for Theoretical Physics, Moscow Center for Advanced Studies, 141701 Moscow, Russia}
\affiliation{National Research Nuclear University MEPhI (Moscow Engineering Physics Institute), Kashirskoe shosse 31, 115409 Moscow, Russia}
\author{J. Cao}%
\affiliation{Abrikosov Center for Theoretical Physics, Moscow Center for Advanced Studies, 141701 Moscow, Russia}
\author{A. Kavokin}
\affiliation{Abrikosov Center for Theoretical Physics, Moscow Center for Advanced Studies, 141701 Moscow, Russia}
\affiliation{Russian Quantum Center, Skolkovo IC, Bolshoy boulevard 30 bld. 1, 121205 Moscow, Russia}
\author{N. Voronova}
\email{nsvoronova@mephi.ru}
\affiliation{National Research Nuclear University MEPhI (Moscow Engineering Physics Institute), Kashirskoe shosse 31, 115409 Moscow, Russia}
\affiliation{Russian Quantum Center, Skolkovo IC, Bolshoy boulevard 30 bld. 1, 121205 Moscow, Russia}
\author{A. Nalitov}
\email{anton.nalitov@gmail.com}
\affiliation{Abrikosov Center for Theoretical Physics, Moscow Center for Advanced Studies, 141701 Moscow, Russia}
\affiliation{Russian Quantum Center, Skolkovo IC, Bolshoy boulevard 30 bld. 1, 121205 Moscow, Russia}

\date{\today}

\begin{abstract}
    In the Supplemental Material, we provide the polariton energy dispersions arising from the linear model, discuss the equilibrium and phenomenologically-modified Gross-Pitaevskii equations, perform stability analysis of the mean-field configurations and fluctuations in the model with gain and saturation, and examine the role of condensate-reservoir interactions. Finally, we derive the fluctuation matrix for the considered fluctuations parameterization 
    \vor{and address the model applicability}. 
\end{abstract}

\maketitle

\section{Polariton dispersions and the equilibrium model}
\label{SMNote1} 
As a starting point of the theory presented in the main text, we introduce the three field operators $\hat P$ and $\hat P_\pm$ of lowest polariton modes. The corresponding branches of the single-particle dispersion are formed due to the coupling between photon waveguide modes and quantum-well excitons. In this Section, we address the formation of the three considered polariton modes within the approach proposed in Ref.~\cite{nigro2025supersolidity}. 

In the linear model of Ref.~\cite{nigro2025supersolidity}, the system can be described by the effective Hamiltonian written in the basis $\left({\rm TE}_{+0}, {\rm TE}_{-0}, {\rm TE}_{+1}, {\rm TE}_{-1}, Q_{+0}, Q_{-0}, Q_{+1}, Q_{-1}\right)^{T}$:
\begin{widetext}
    \begin{equation}\label{hamiltonian}
    \hat{H}_0 = \begin{pmatrix} \hbar \omega_{+0} - i \gamma_{\rm C} & U +  i  \gamma_{\rm C} & 0 & 0 & \frac{\hbar \Omega_0}{2} & 0& 0& 0 \\
   U +  i  \gamma_{\rm C} & \hbar \omega_{-0} - i \gamma_{\rm C} & 0 & 0& 0 &  \frac{\hbar \Omega_0}{2}  & 0 & 0 \\
   0 & 0&  \hbar \omega_{+1} - i \gamma_{\rm C} & 0 & 0 & 0 & \frac{\hbar \Omega_1}{2} & 0 \\
   0 & 0& 0 & \hbar \omega_{-1} - i \gamma_{\rm C} & 0 & 0 & 0 & \frac{\hbar \Omega_1}{2}\\
    \frac{\hbar \Omega_0}{2} & 0 & 0 & 0 & E_{\rm X} & 0 & 0 & 0 \\
    0 &  \frac{\hbar \Omega_0}{2} & 0 & 0 & 0 & E_{\rm X} & 0 & 0\\
    0 & 0 &  \frac{\hbar \Omega_1}{2} & 0 & 0 & 0 &  E_{\rm X} & 0 \\
    0 & 0 & 0 &  \frac{\hbar \Omega_1}{2} & 0 & 0 & 0 & E_{\rm X}  
    \end{pmatrix} 
    \end{equation}
\end{widetext}
where 
$E_{\rm X}({\bf k}) = E_{\rm X}^0 + \frac{\hbar^2 k^2}{2 m_{\rm X}} - i\gamma_{\rm X}$
is the exciton energy dispersion 
(with non-radiative losses $\gamma_{\rm X}$) corresponding to the exciton modes $Q_i$ with the effective mass $m_{\rm X}$,  $\hbar \Omega_{0(1)}$ is the Rabi splitting between the exciton $Q_{\pm 0(1)}$ and ${\rm TE}_{\pm 0(1)}$ photon modes, 
$$\hbar \omega_{\pm i}({\bf k})\approx \hbar \omega^{0}_{i} \pm \frac{\hbar c k_x}{n_{\rm g}} + \frac{\hbar c a }{4 \pi n_{\rm g}}k_y^2$$ 
describes the $(\pm i)$ optical waveguide mode with radiative losses $\gamma_{\rm C}$, and $U$ is the diffractive coupling constant. The energy dispersion laws of photons depend on the geometry of the sample, i.e., on the grating period $a$ and the refractive index $n_{\rm g}$. In the model, $+0$ and $-0$ modes are coupled to each other due to the diffraction mechanism and the radiative coupling. We note that every photon mode ${\rm TE}_{i}$ is coupled to  
a separate exciton mode $Q_i$. We neglect here the coupling between the ${\rm TE}_{\pm0}$ and ${\rm TE}_{\mp1}$ modes since in the range of considered energies and momenta, the difference between dispersions obtained when taking into account coupling between ${\rm TE}_{\pm0}$ and ${\rm TE}_{\mp1}$ modes and dispersions in the absence of the coupling is negligible. 
    
The diagonalization of the $8\times 8$ Hamiltonian~(\ref{hamiltonian}) yields the new eight polariton modes. We are interested in the three modes taking part in parametric scattering: the negative effective mass mode supporting the BiC condensate, arising due to the coupling of excitons to $\pm 0$ photons, and the two adjacent branches originating from the strong coupling between the excitons and photonic ${\rm TE}_{\pm1}$ modes. Thus we can reduce the model and consider the $3\times3$ Hamiltonian. 

\begin{figure*}[t!]
    \centering
\includegraphics[width=1\linewidth]{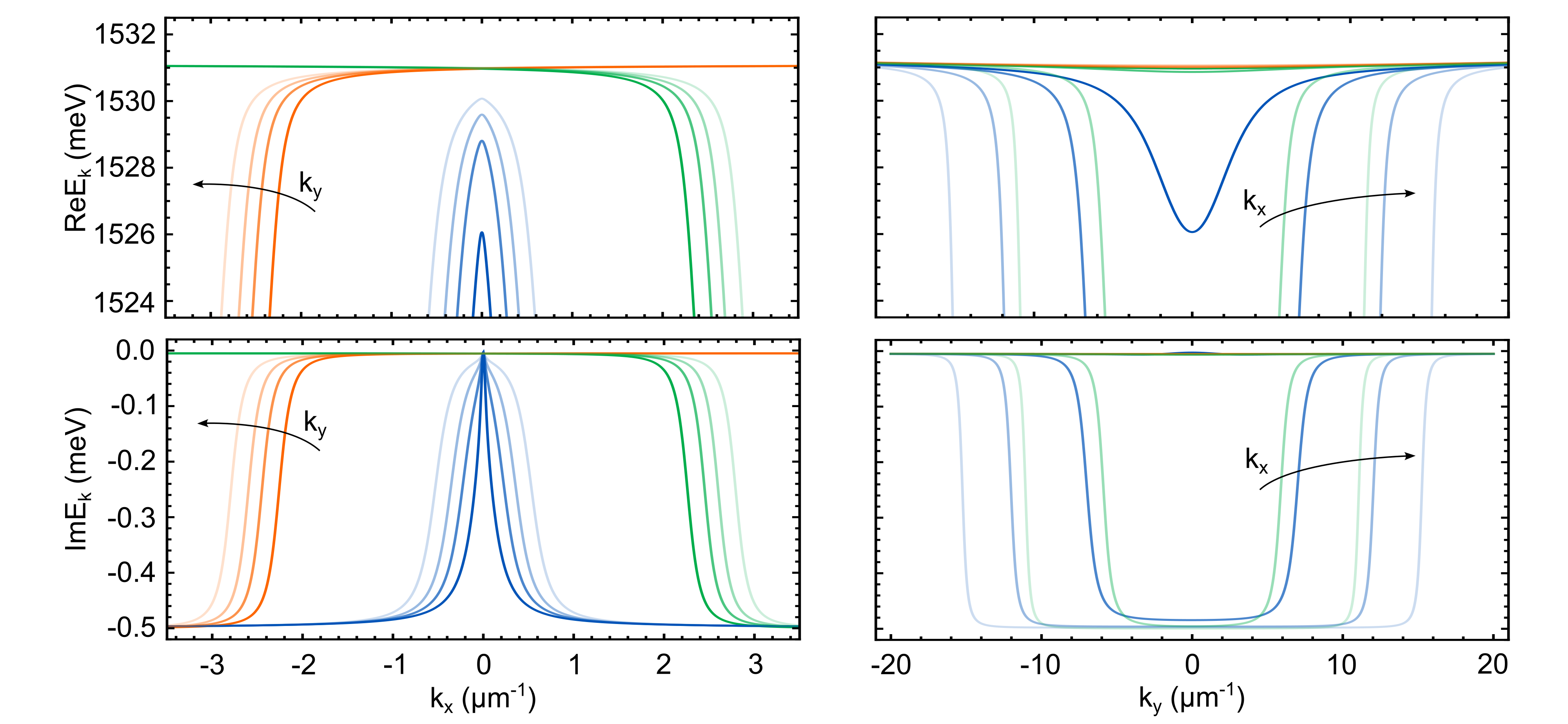}
    \caption{Real (top) and imaginary (bottom) parts of the polariton energy dispersions for the main (BiC) mode $\varepsilon({\bf k})$ (the blue lines) and the adjacent modes $E_{+}({\bf k})$ (the orange lines) and $E_{-}({\bf k})$ (the green lines) from the $3\times3$ model. Left column: against $k_x$ at     $k_y = 0$, 3, 4, 5~$\mu$m$^{-1}$ (from darker to lighter lines shading). Right column: 
    against $k_y$ at 
    $k_x = 0$, 1, 3, 5~$\mu$m$^{-1}$ (from darker to lighter lines shading). Parameters: $n_{\rm g} = 3.34$, $a=242$~nm, $U = -1.9$~meV, $\hbar\Omega_0 = 11.2$~meV, $\hbar\Omega_1 = 8.4$~meV, $E_{\rm X}^0 = 1531.1$~meV, $m_{\rm X}=0.22~m_0$,  $\hbar\omega_0^0=1533.58$~meV, $\gamma_{\rm C} = 0.5$~meV, $\gamma_{X}= 10^{-2} \gamma$, $\hbar \omega_{\pm 1}^0 - \hbar \omega_{\pm0}^0=130$~meV.}
    \label{SFig1}
\end{figure*}

We use the expression for the lower lower (LLP) branch previously derived 
in~\cite{ardizzone2022polariton}, 
resulting from the coupling of the antisymmetric photon mode 
$$E_{\rm C}({\bf k}) = \hbar \omega^0_0 \!-\! \vor{i\gamma_{\rm C}} \!+\! \frac{\hbar c a }{4 \pi n_{\rm g}}k_y^2 \!-\! \sqrt{\left(\frac{\hbar c k_x}{n_{\rm g}}\right)^{\!\!2} \!+\! \left(U \!+\! \vor{i  \gamma_{\rm C}} \right)^2}$$ 
with corresponding exciton mode $Q$. The energy dispersion 
of the LLP 
(BiC)  mode thus reads: 
$$\varepsilon({\bf k}) = \frac{E_{\rm C} + E_{\rm X}}{2} - \frac{1}{2}\sqrt{\left(E_{\rm C} - E_{\rm X}\right)^2 + (\hbar \Omega_0)^2}\,.$$ 
The adjacent branches dispersions are then easily obtained from Eq.~(\ref{hamiltonian}): 
$$E_{\pm}\!({\bf k}) = \frac{\hbar \omega_{\pm 1} \!-\! i \gamma_{\rm C} \!+\! E_{\rm X}}{2} \!-\! \frac{\sqrt{(\hbar \Omega_1)^2 \!+\! (\hbar \omega_{\pm1} \!-\! i\gamma_{\rm C} \!-\! E_{\rm X})^2}}{2}.$$
The exciton Hopfield coefficients for the three derived modes are as follows:
\begin{subequations}
\begin{align}
    & X_{\bf k} = \frac{1}{\sqrt{2}}\biggl(1 + \frac{(E_{\rm C} - E_{\rm X})}{\sqrt{(E_{\rm C} - E_{\rm X})^2+ (\hbar\Omega_0)^2}}\biggr)^{\!\!1/2}\!\!\!\!\!\!, \nonumber \\
    & X_{\pm1}({\bf k}) = \frac{1}{\sqrt{2}}\biggl(1 + \frac{\hbar\omega_{\pm1}-i\gamma_{\rm C} - E_{\rm X}}{\sqrt{(\hbar \omega_{\pm1} -i\gamma_{\rm C} - E_{\rm X})^2 + (\hbar \Omega_1)^2}}\biggr)^{\!\!1/2}\!\!\!\!\!\!.\nonumber
\end{align}
\end{subequations}

The obtained energy dispersions $\varepsilon({\bf k})$ and $E_{\pm}\!({\bf k})$, both the real and imaginary parts, are plotted in Fig.~\ref{SFig1} for the parameters of the sample used in the experiment~\cite{trypogeorgos2025emerging}, against $k_x$ at different fixed $k_y$ and against $k_y$ at different fixed $k_x$. Importantly, as can be seen from the left-hand side of the figure, the main mode  $\varepsilon({\bf k})$ is characterized by the negative mass in the $k_x$ direction, and the point ${\bf k}=0$ corresponds to the BiC state with zero losses. This is the presence of the negative-mass dispersion that results in the instabilities discussed below. Note that in the following, only one dimension (along $k_x$ at $k_y=0$) is considered, see discussion in the main text.

Once we have refined our approach to consider only the three lowest modes, it becomes straightforward to define the nonlinear processes, including the optical parametric scattering. 
%
Using the second-quantized Hamiltonian (1) of the main text, we derive the Heisenberg equations for the polariton field operators $\hat P({\bf r},t)$ and $\hat P_{\pm1}({\bf r},t)$: $i\hbar\partial_t\hat P_i = \bigl[\hat P_i,\hat H-\mu\hat N\bigr]$, where $\hat N$ is the particle number operator and $\mu$ the chemical potential. They read:
\begin{widetext}
    \begin{subequations}\label{A2}
    \begin{align}
    &\!\!\! i\hbar  \frac{\p}{\p t}\hat{P}({\bf r},t)\!=\!\bigl(\varepsilon(\hat{\bf k})-\mu\bigr)\hat{P}({\bf r}, t) \!+\!g \!\! \int \!\!d{\bf r'} X^*\!({\bf r}'\!\!-\! {\bf r}) \hat{Q}^{\dag}({\bf r}'\!, t)\hat{Q}({\bf r}'\!, t)\hat{Q}({\bf r}'\!, t) \!+\! 2 \tilde{g} \!\!\int\!\! d{\bf r}' X^*\!({\bf r}'\!\! -\! {\bf r}) \hat{Q}^{\dag}({\bf r}'\!, t)\hat{Q}_{+}\!({\bf r}'\!, t)\hat{Q}_{-}({\bf r}'\!, t),\\
      &\!\!\! i\hbar\frac{\p}{\p t}\hat{P}_{\pm 1}({\bf r}, t)= \bigl(E_{\pm 1}(\hat{\bf k})-\mu\bigr) \hat{P}_{\pm 1}({\bf r},t) + \tilde{g} \!\!\int\!\! d{\bf r}' X_{\pm}^*({\bf r}'\!\! -\! {\bf r}) \hat{Q}_{\mp}^{\dag}({\bf r}'\!, t)\hat{Q}^2({\bf r}'\!, t).
\end{align}
\end{subequations}
\end{widetext}
Here $\hat{\bf k}=-i\nabla$, $X({\bf r})$, $X_\pm({\bf r})$ are the inverse Fourier images of $X_{\bf k}$ and $X_{\pm }({\bf k})$, respectively; the interaction constants $g$, $\tilde g$ and connection between the field operators $\hat Q_i$ and $\hat P_i$  are introduced in the main text.

Given 
that the macroscopic populations arise on the BiC branch at the saddle point (at $k=0$) and on the adjacent 
branches at $\pm {\bf k}_0$, 
we can employ the standard (equilibrium) mean-field approach and introduce the order parameter components as $\langle \hat{P}({\bf r}, t) \rangle  = \Psi_0({\bf r})$ and $\langle \hat{P}_{\pm 1}({\bf r}, t) \rangle  = \Psi_{\pm}({\bf r})$. 
Averaging of Eqs.~\eqref{A2} leads to the coupled Gross-Pitaesvkii Equations (GPEs) for the macroscopic wavefunctions $\Psi_0$ and $\Psi_{\pm}$.
Endowing them the time-dependence 
$\Psi_i({\bf r},t) = \Psi_i({\bf r})e^{-i\mu t/\hbar}$, where $\mu$ has the meaning of the condensate energy, we get:
     \begin{align}\label{A3}
     i\hbar \frac{\partial}{\partial t}\Psi_{0}({\bf r}, t) &=\! \varepsilon(\hat{\bf k})\Psi_{0}({\bf r}, t) + g |X_0|^4 |\Psi_{0}({\bf r}, t)|^2 \Psi_{0}({\bf r}, t)  \nonumber\\ 
     & \quad + 2 \tilde{g} \tilde{X} \Psi_{0}^*({\bf r}, t) \Psi_{-}({\bf r}, t)\Psi_{+}({\bf r}, t) ,\\
     i \hbar \frac{\partial}{\partial t}\Psi_{\pm}({\bf r}, t) &=\! E_{\pm}(\hat{\bf k})\Psi_{\pm}({\bf r}, t) \!+\! \tilde{g} \tilde{X}^{*} \Psi_{\mp}^*({\bf r}, t) \Psi_{0}^2({\bf r}, t) \nonumber
\end{align}
with $\tilde{X} = |\tilde{X}|e^{i \delta \phi} 
= (X_{0}^*)^2 X_{+1}(-{\bf k_0})X_{-1}({\bf k_0})$. 

\section{Phenomenological modification of Gross-Pitaevskii Equations}
\label{SMNote2-}

In this Section, we address the stability of the GPEs obtained in Sec.~\ref{SMNote1} when they are modified to account for the continuous nonresonant pumping.

The first equation of the GPEs~\eqref{A3} 
can be phenomenologically coupled to the incoherent exciton reservoir~\cite{PhysRevLett.99.140402}, namely,
\begin{subequations}
     \begin{align}
     i\hbar \frac{\partial\Psi_{0}}{\partial t} & \!\!=\! \varepsilon(\hat{\bf k})\Psi_{0} \!\!+\! g |X_0|^4 |\Psi_{0}|^2 \Psi_{0} \!\!+\! 2 \tilde{g} \tilde{X} \Psi_{0}^* \Psi_{\!-}\Psi_{\!+} \!\!+\! iRn_{\!R}\Psi_0, \nonumber\\ 
     \frac{\p n_R}{\p t}\,\, & \!\! = \mathcal{W} - \gamma_R n_R - R |\Psi_0|^2 n_R, \nonumber
\end{align}
\end{subequations}
where $n_R$ denotes the reservoir density, $\mathcal{W}$ is the intensity of the non-resonant pumping, $\gamma_R$ is the reservoir loss rate, and $R$ governs the rate of  stimulated scattering from the reservoir into the main (BiC) mode. Note that the condensate linear losses are implicitly included since $\varepsilon({\bf k})$ is complex. 

When the reservoir is adiabatically eliminated in a standard way~\cite{bobrovska,berloff2017polariton}: 
$n_R \approx \frac{\mathcal{W}}{\gamma_R}\Bigl(1 - \frac{R}{\gamma_R}|\Psi_0|^2\Bigr)$, 
this results in the appearance of the effective pumping of the power $W= \mathcal{W} R/\gamma_R$ and gain saturation with the characteristic rate  $\eta = R/\gamma_R$ in the GPEs~\eqref{A3}:
\begin{subequations}\label{gpe_pump}
\begin{align}
     i\hbar \frac{\partial}{\partial t}\Psi_{0} & = \! \varepsilon(\hat{\bf k})\Psi_{0} + (g |X_0|^4 -i \eta W)|\Psi_{0}|^2 \Psi_{0}  \nonumber\\ 
     & \qquad\qquad\,\, + 2 \tilde{g} \tilde{X} \Psi_{0}^* \Psi_{-}\Psi_{+} + i W \Psi_0,  \label{gpe_pump1}\\
     i \hbar \frac{\partial}{\partial t}\Psi_{\pm}&= \! E_{\pm}(\hat{\bf k})\Psi_{\pm} \!+\! \tilde{g} \tilde{X}^{*} \Psi_{\mp}^* \Psi_{0}^2.
\end{align}
\end{subequations}
The parameters entering the equations, $W$ and $\eta$, implicitly depend on the reservoir loss and scattering rates $\gamma_R$ and $R$, 
whose values are usually taken from fitting 
and depend on the sample and experimental configurations. The existing literature reports a wide range for both parameters, in particular,  $\gamma_R$ from $0.6$ to $330~\mu$eV and $R$ from approx. $0.1$ to $200~\mu$eV~$\mu$m$^2$, see e.g. Refs.~\cite{PhysRevLett.123.047401,PhysRevB.91.085413,tosi2012sculpting,comaron}.

\vor{It is important to note that, strictly speaking, the gain-saturation model employed here has a limited applicability in terms of the condensate density $n_0=|\Psi_0|^2$. For pump powers considered in this work, the parameter of expansion $\eta n_0$ varies between $\sim 0.17$ in the case of Fig.~2(b) up to $\sim0.7$ in Fig.~3 of the main text. While the latter value seems to be out of the validity range of the model, below we check that retaining the full shape of the reservoir density $n_R = W/[R(1-\eta|\Psi_0|^2)]$ in Eqs.~(2) does not qualitatively affect the reported physics.}

\vspace{-5pt}
\subsection{Stationary configurations}

\vspace{-2pt}
Prior to addressing fluctuations above the order parameters components, we first check the stability of the mean-field stationary configurations.
Parameterizing the order parameters in terms of the phase and density as given in Eqs.~(3) of the main text,
\vor{and approximating the dispersions
\begin{equation}\label{E_approx}
\begin{split}
& {\rm Re}\,\varepsilon(\hat{\bf k})  \simeq -\hbar^2\partial_{xx}^2/2m,  \\
&{\rm Re}E_\pm(\hat{\bf k})  \simeq {\rm Re} E_{\pm}(\mp {\bf k}_0) \mp i \hbar v \partial_x, 
\end{split}
\end{equation}
where $m$ is the (negative) effective mass of the BiC mode in the $k_x$-direction and $\pm v$ is the linear slope coefficient of the $\pm1$ modes at $k_x \sim \mp k_0$, from~\eqref{gpe_pump}}
one gets: 
\begin{widetext}
    \begin{subequations}\label{ra_eq}
    \begin{eqnarray}
   \hbar \frac{\partial n_0}{\partial t} &=& \Bigl(\!\vor{-\frac{\hbar^2\partial_{xx}^2}{m}\phi_0} + 2 {\rm Im}\varepsilon(0) + 2 W - 2 {\rm Im \mu} \!\Bigr)n_0 - 2 \eta W n_0^2+ 4\tilde{g} |\tilde{X}| n_0 \vor{\sqrt{n_+n_-}} \sin{(\delta \phi + \phi_+ + \phi_- \!\vor{-2\phi_0})}, \label{ra_eq1}\\
   \hbar \frac{\partial n_{\vor{\pm}}}{\partial t} &=& (\vor{\mp\hbar v\partial_x} + 2 {\rm Im}E_{\pm}(\mp {\bf k}_0)- 2 {\rm Im \mu}) n_{\vor{\pm}} - 2 \tilde{g}|\tilde{X}| n_0  \vor{\sqrt{n_+n_-}} \sin{(\delta \phi + \phi_+ + \phi_-\!\vor{-2\phi_0})},\label{ra_eq2}\\
   \vor{\hbar \frac{\partial \phi_0}{\partial t}} &\vor{=}& \vor{\frac{\hbar^2(\partial_x n_0)^2}{8mn_0^2} + \frac{\hbar^2\partial_{xx}^2n_0}{4mn_0} - \frac{\hbar^2(\partial_x
   \phi_0)^2}{2m}+ {\rm Re}\mu - g|X_0|^4n_0 - 2 \tilde{g} |\tilde{X}| \sqrt{n_+n_-}\cos{(\delta \phi + \phi_+ + \phi_-\!\vor{-2\phi_0})},}\label{ra_eq_phi0}\\
   \hbar \frac{\partial \phi_{\pm}}{\partial t} &=& -{\rm Re} E_{\pm}(\mp {\bf k}_0) \vor{\,\,\mp\,\, \hbar v\partial_x\phi_\pm} + {\rm Re}\mu - \tilde{g} |\tilde{X}| n_0\cos{(\delta \phi + \phi_+ + \phi_-\!\vor{-2\phi_0})}.\label{ra_eq3}
\end{eqnarray}
\end{subequations}
\end{widetext}
In the main text we introduced the notation $|{\rm Im}\varepsilon(0)| \equiv \gamma$ for brevity.
\vor{Without the loss of generality, one can set the stationary reference phase $\phi_0=0$; 
the stationary uniform mean-field values of $n_0$ and $\langle n_\pm\rangle = n_1$ are defined below.}
The equation for the phases $\phi_{\pm}$~(\ref{ra_eq3}) shows that \vor{for the stationary configuration,} in the case $\cos(\delta\phi+\phi_+ + \phi_-)\ne 0$, particles from the condensate scatter into the states with momenta $\pm {\bf k}$ which are not lying at the same energy with the condensate, and thus this process is breaking the energy conservation law. Hence, we require $\cos{(\delta\phi+\phi_+ + \phi_-)}= 0$, which means that the OPO-process leads to the leakage from the BiC mode and gain of the $\pm1$ modes. This corresponds to the phase relation $\delta\phi+\phi_+ + \phi_- = -\pi/2$, and the scattering momenta $\pm{\bf k}_0$ are then defined from ${\rm Re}\mu(W) = {\rm Re}E_{\pm}(\mp {\bf k}_0)$.
Note that while from Eq.~\eqref{ra_eq_phi0} it follows that the condensate energy $\mu= g|X_0|^4 n_0$ is not dependent on the pump power, the final model in Eq.~(2) of the main text yields the dependence: 
$$\mu(W) = \Bigl(g |X_0|^4 -g_R \frac{W}{R}\Bigr)n_0 + g_R \frac{W}{\vor{R}}.$$ 
In the expressions above, we assume that $\mu$ is real~\cite{footnote} and that losses from the adjacent $\pm1$ modes are momentum-independent (see discussion in the main text): 
${\rm Im}E_{\pm}({\mp {\bf k}}_0)\approx {\rm const}$. 

It is useful to define the threshold pump values as well as the dependencies of the densities on pumping. From Eqs.~\eqref{ra_eq1}--\eqref{ra_eq2}, the first and second threshold pump powers 
are defined as 
\begin{align}
W_{\rm th1} &= |{\rm Im} \varepsilon(0) - {\rm Im}\mu|,  \label{th1}  \\
W_{\rm th2} &= \tilde{g}|\tilde{X}||{\rm Im} \varepsilon(0)|/(\tilde{g}|\tilde{X}| - \eta |{\rm Im} E_{\pm}(\mp {\bf k}_0)|).  \label{th2}
\end{align}
We note that the second threshold value $W_{\rm th2}$ depends on experimental sample parameters via $\eta=R/\gamma_R$ and on the detuning (exciton fractions) via $\tilde X$. Hence, when plotting pump-power dependent results (e.g. in Figs.~\ref{SFig2},~\ref{SFig4}(a) and the inset of Fig.~2 in the main text), we normalize the values of $W$ to a fixed number $W_0$ that corresponds to $W_{\rm th2}(|\tilde X|^2=0.226,\eta =1.4 \times10^{-3}$~$\mu$m$^{-2})$.

The stationary densities 
above $W_{\rm th2}$ are as follows:
\begin{subequations}\label{stationary}
    \begin{align}
       &  n_0 = \frac{|{\rm Im} E_{\pm}(\mp {\bf k}_0) - {\rm Im}\mu| }{\tilde{g}|\tilde{X}|},\\
       & n_1(W) = \frac{{\rm Im} \varepsilon(0) - {\rm Im}\mu + W}{2 \tilde{g}| \tilde{X}|} - \frac{\eta W n_0}{2\tilde{g}|\tilde{X}| }.
    \end{align}
\end{subequations}

\subsection{Stability analysis}

\begin{figure}[t!]
    \centering
\includegraphics[width=\linewidth]{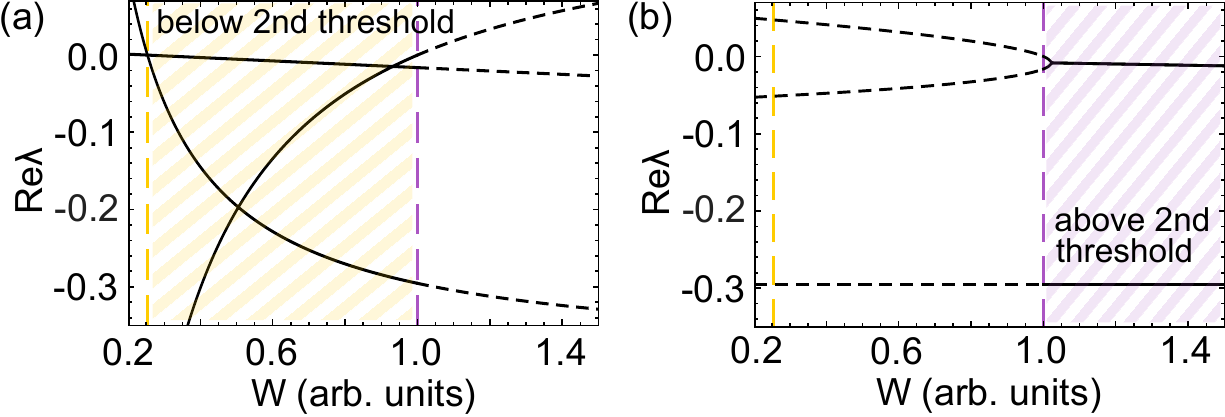}
    \caption{The real parts of the Lyapunov exponents ${\rm Re \lambda}$ for the stationary configurations of Eqs.~\eqref{ra_eq}. (a) 
    Below the second threshold, $n_0 = ({\rm Im}\varepsilon(0) + W - {\rm Im}\mu)/(\eta W)$, $n_1=0$, $\phi_\pm=0$. (b) above the second threshold, see Eq.~\eqref{stationary}. The yellow and lilac dashed vertical lines indicate the first and second thresholds according to Eqs.~\eqref{th1},~\eqref{th2}, respectively. The stationary configurations both below and above $W_{\rm th2}$ result in ${\rm Re\lambda}<0$.}
    \label{SFig2}
\end{figure}

The stability analysis of the equations~\eqref{ra_eq} shows that the Lyapunov exponents both below and above the second threshold have negative real parts (see Fig.~\ref{SFig2}), so that the stationary configurations for $n_0$, $n_{1}$, and $\phi_{\pm}$ can be assumed to be stable. However, let us note that while the negative real values of the Lyapunov exponents are necessary, they are not sufficient for stability of the order-parameter components: 
as we will see below, unstable fluctuations destroy the order parameter. 


\begin{figure*}[t]
\includegraphics[width=1\textwidth]{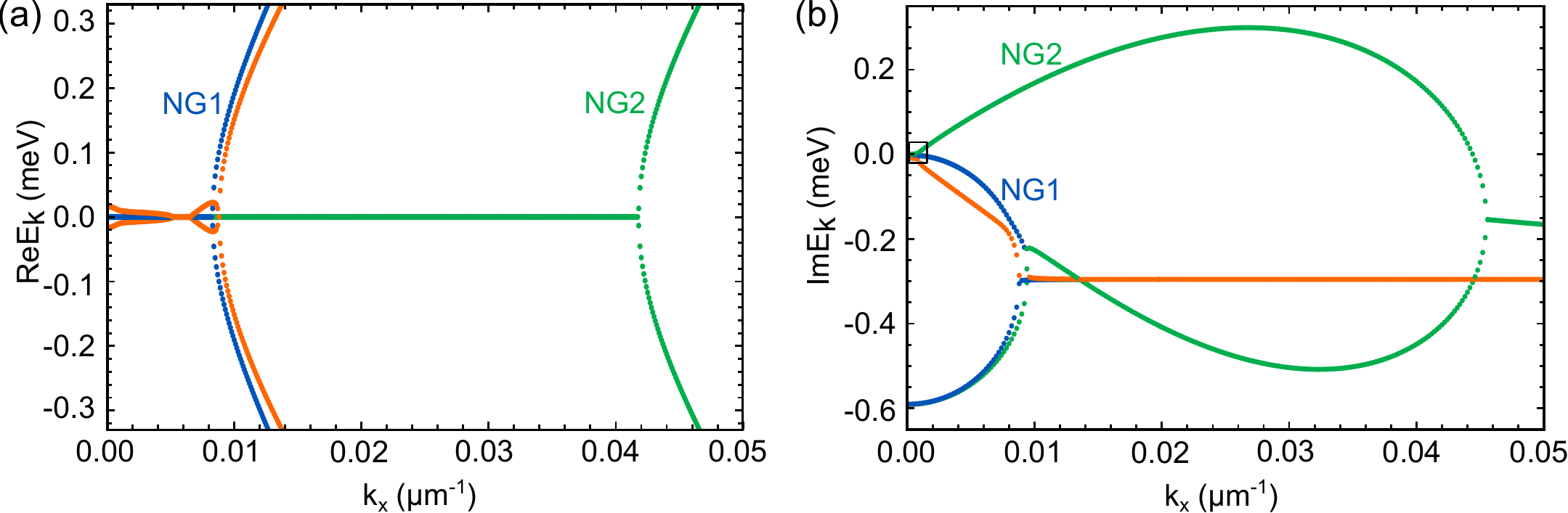}
    \captionsetup{width=\textwidth} 
    \caption{The real (a) and imaginary (b) parts of the first band of the spectrum $E_{\bf k}$ of elementary excitations on top of the mean-field stationary solution~\eqref{stationary} calculated within the model~\eqref{gpe_pump} without the condensate-reservoir interactions, as a function of $k_x$ at $k_y=0$. \vor{The small rectangle at $k_x\to0$ in panel (b) indicates the region plotted in Fig.~4(b) of the main text.}
    Parameters: $W = 1.1W_{\rm th2}$, $n_0 = 456~\mu$m$^{-2}$, $n_1/n_0 = 5\times 10^{-4}$, $\mu = 0.35$~meV, $g = \tilde g = 2.5~\mu$eV~$\mu$m$^{2}$, $\gamma_R = 4.8~\mu$eV, $R =6.7 \times10^{-3}~\mu$eV~$\mu$m$^{-2}$. Other parameters as in Fig.~\ref{SFig1}. }
    \label{SFig3}
\end{figure*}

We calculate the spectrum of elementary excitations on top of the stationary configuration of Eqs.~\eqref{ra_eq} above the second threshold [see Eqs.~\eqref{stationary}] using the same ansatz for fluctuations as in Eqs.~(6) of the main text. The resulting spectrum, plotted in Fig.~\ref{SFig3}, \vor{reveals at instability}. 
More precisely, the first band of the excitation spectrum consists of six branches: ({\it i}) the gapped pair with the negative imaginary part (the orange lines in Fig.~\ref{SFig3}),  ({\it ii}) two modes with  ${\rm Re} E_{\bf k} \to0$ and nonzero ${\rm Im}E_{\bf k} < 0$, and ({\it iii}) two gapless modes corresponding to $ E_{\bf k} \to0$. 
%
\vor{As shown analytically in the End Matter of the main text,} in terms of the phase and amplitude fluctuations, the two gapless modes (one possesses positive imaginary part, while the other has negative one at $k\ne 0$) \vor{in the limit $k\to0$} correspond to phase fluctuations, 
and 
can be addressed as Nambu-Goldstone modes. At the same time, \vor{here---for the conventional case of Eqs.~\eqref{gpe_pump} with positive $g$}---one sees that at finite $k$, one of these NG modes 
(the green line in Fig.~\ref{SFig3}) 
is gained, \vor{the wavevectors corresponding to unstable perturbations are not confined to the region of $k\to0$,} and that phase fluctuations destroy the long-range order in the supersolid state of the system. 
\subsection{Condensate-reservoir interaction as a stabilization mechanism}\label{SMNote3}

Finally, based on the above analysis we arrive at the necessity to modify the model and take into account condensate-reservoir interactions, by adding the term $\propto g_R n_R\Psi_0$ to the GPE~\eqref{gpe_pump1} for the condensate macroscopic wavefunction. Such modifications have been previously considered as additional blueshift~\cite{PhysRevLett.99.140402,Sigurdsson2024,comaron} with the aim to achieve a better correspondence of simulations results with the experiment. Here, the inclusion of the condensate-reservoir interactions analytically proves to be crucial for the observed phases stability.

Given the adiabatic approximation for the reservoir density \vor{expanded to first order} $n_R \approx \frac{W}{R}\bigl(1 - \frac{R}{\gamma_R}|\Psi_0|^2\bigr)$ as discussed in the beginning of SM~Sec.~\ref{SMNote2-}, one finally arrives at the Eqs.~(2) of the main text: apart from the additional constant blueshift $g_R\frac{W}{R}$, the GPE acquires the renormalized interaction $g_{\rm eff}(W)|\Psi_0|^2\Psi_0$ with
\vspace{-2pt}
\begin{equation}
g_{\rm eff}(W) = g|X_0|^4 - g_R\frac{W}{\gamma_R}.
\vspace{-3pt}
\end{equation}
In the main text, we emphasize that the effective interaction constant $g_{\rm eff}(W)$ depending on the exciton fraction $|X_0|^2$ and the pump power $W$ may be negative-valued, which leads effectively to the stabilization of the system \vor{both below and above the second threshold}.

\begin{figure*}[t]
\includegraphics[width=1\linewidth]{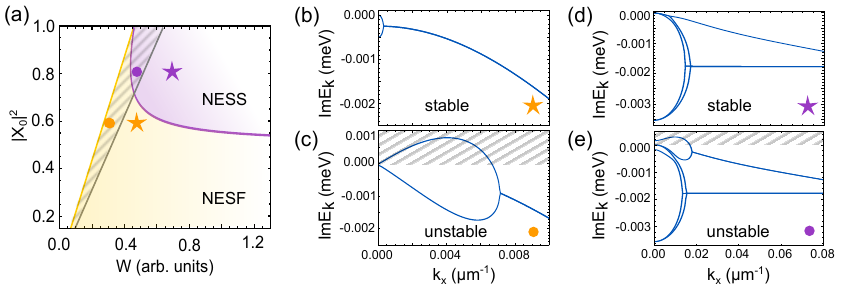}
    \caption{(a) The phase diagram calculated using the Eq.~(2) of the main text, for $R=0.01~\mu$eV$~\mu$m$^{2}$ and $\gamma_R=0.007~\mu$eV. The yellow and purple lines indicate the boundaries for the condensed  (NESF) and supersolid (NESS) phases. The gray line shows the stabilization boundary. The hatched region shows the unstable region. Panels (b)--(e) show the imaginary parts of the excitation spectrum vs. $k_x$ corresponding to different phases as marked: (b) $W=1.14W_{\rm th2}$, (c) $W=0.68W_{\rm th2}$, (d) $W=1.76W_{\rm th2}$, (e) $W=0.88W_{\rm th2}$. Interaction constant $g = \tilde g = 2.5~\mu$eV~$\mu$m$^{2}$, other parameters as in Fig.~\ref{SFig1}.}
    \label{SFig4}
\end{figure*}

To summarize the role of the reservoir interaction in the stabilization of elementary excitations, here we calculate the phase diagram that is to be compared to the inset of Fig.~2(a) of the main text, choosing another value of the reservoir dissipation and scattering rates $\gamma_R$, $R$. In this case, the stabilization threshold related to the sign-reversal of the interaction constant happens at the pump powers {\it above} the first (condensation) threshold ($W>W_{\rm th1}$), see Fig.~\ref{SFig4}. This fact leads to arising of the new phases, i.e., the unstable NESS and unstable NESF, which are shown in Fig.~\ref{SFig4}(a),~(c) and~(e) as 
hatched areas. As anticipated, the corresponding excitation spectra demonstrate the positive imaginary parts (see panels indicated with circles in Fig.~\ref{SFig4}) which leads to quenching of supersolidity in the thermodynamic limit due to growing fluctuations. 

In the calculations of the phase diagrams in Fig.~\ref{SFig4} and Fig.~2 of the main text, we changed the Hopfield coefficient $|X_0|^2$ introducing the detuning $\Delta$ of the ${\rm TE}_{\pm0}$ photonic modes with respect to the exciton energy $E_{\rm X}^0$, 
which can be achieved by slightly tuning the grating period of the structure $a$ along the waveguide direction. In particular, the excitation spectra in Fig.~\ref{SFig4} are calculated using $\Delta = 3.5$~meV for 
Fig.~\ref{SFig4}(b),~(d) and 10.5~meV for the 
Fig.~\ref{SFig4}(c),~(e). In all calculations presented in the main text, the detuning was taken $\Delta = 2.5$~meV (corresponding to the exciton fraction $|X_0|^2=0.55$, marked by the dashed line in the phase diagram shown in the inset of Fig.~2).

{\section{Fluctuation matrix}
\label{SMNote3-}}
In this work, we consider fluctuations within the parametrization using the Bloch theorem (see Eqs.~(6) of the main text). In this supplementary Section, the fluctuation matrix is derived. 

In the basis of the Bogoliubov amplitudes $\bigl(u_{l, k+m k_1}\bigr.$,  $v_{l, k+m k_1}$,  $u^{(+)}_{l, k+m k_1}$, $v^{(+)}_{l, k+m k_1}$, $u^{(-)}_{l, k+m k_1}$, $\bigl.v^{(-)}_{l, k+m k_1}\bigr)^{T}$, where $l$ is the number of a band, $k_1 = k_0(W)$ is the periodicity wavevector, $m\in\mathbb{Z}$  ($u_{l,k},\,v_{l,k}$ are the Bogoliubov amplitudes for the main mode, while $u_{l,k}^{(\pm)},\, v_{l,k}^{(\pm)}$ 
are the Bogoliubov amplitudes for the adjacent $\pm1$ modes), the fluctuation matrix has, in fact, infinite dimension. In our work, however, we focus on the range of small momenta, i.e. $k\ll k_0$, so all the calculations are 
truncated for the treatment of the first energy band. In this regard, it is reasonable to approximate the full infinite-dimensional matrix to the consideration of $m=\overline{-2,2}$. 

The fluctuation matrix is derived in the standard fashion and can be expressed in the block shape
   \begin{equation} \label{S12}
   \hat{M} = 
        \begin{pmatrix}
    \dots & \hat{M}^{a} & \hat{M}^{d}_{m-1} & \hat{M}^{b} & 0 & 0&\dots\\[3pt]
    \dots & 0 & \hat{M}^{a} & \hat{M}^{d}_{m} & \hat{M}^{b} & 0 & \dots \\[3pt]
    \dots & 0 & 0 &\hat{M}^{a} & \hat{M}^{d}_{m+1} & \hat{M}^{b}  & \dots 
\end{pmatrix},
   \end{equation}
where $\hat{M}^{d}_{m} \equiv \hat{M}^{d}({\bf k} + m {\bf k}_0)$ is the $6\times6$ matrix on the main diagonal, which is dependent on momentum:
\begin{widetext}
    \begin{equation}\label{matrix1}
    \hat{M}^d({\bf k})=
    \begin{pmatrix}
        M_{11}({\bf k}) & M_{12} &0 & 0 & 0 & 0 \\ 
        -M^{*}_{12} & -M^*_{11}(-{\bf k}) & 0 & 0 & 0 & 0 \\
        0 & 0 & M_{33}({\bf k}) & 0 & 0 & M_{36} \\
        0 & 0 & 0 & -M^{*}_{33}(-{\bf k}) & M_{36} & 0 \\
        0 & 0 & 0 &  M_{36} & M_{55}({\bf k}) & 0 \\
        0& 0 & M_{36} & 0 & 0 & - M^{*}_{55} (-{\bf k})
    \end{pmatrix},
\end{equation}
$\hat{M}^{a(b)}$ is the $6\times6$ matrix:
\begin{equation}\label{matr2}
\hat{M}^{a}=
\begin{pmatrix} 
    0 & 0 & M_{13} & 0\,\, &0 & 0 \\ 
        0 &0& 0 & 0\,\, & 0 & M_{13} \\
        0 & 0 &0 & 0\,\, & 0 & 0 \\
        0 & M^{*}_{13} & 0 & 0\,\, & 0 & 0 \\
        M^{*}_{13} & 0 & 0 & 0\,\, & 0 & 0 \\
        0&  0 &0 & 0\,\, & 0 &0 
    \end{pmatrix}, \quad \hat{M}^{b}=
\begin{pmatrix} 
   0 &0 & 0& 0 & M_{13} & 0\,\, \\ 
       0 &0& 0 & M_{13} & 0 & 0\,\, \\
        M^{*}_{13} & 0 &0 & 0 & 0 & 0\,\, \\
        0 &0 & 0 & 0&0 & 0\,\, \\
        0& 0 & 0 &  0 & 0& 0\,\, \\
        0&  M^{*}_{13} & 0 & 0 & 0 &0 \,\,
    \end{pmatrix}.
\end{equation}
In Eqs.~(\ref{matrix1}--\ref{matr2}), the following notations are introduced:
\begin{subequations}
    \begin{align*}
       & M_{11}({\bf k}) =\varepsilon({\bf k}) - \mu + i W + g_R \frac{W}{R} + 2 (g_{\rm eff}(W) - i \eta W) n_0,\quad M_{12} = -(g_{\rm eff} (W)) - i \eta W) n_0 + 2i \tilde{g}|\tilde X| n_1,\\
       & M_{13} = - 2 i \tilde{g}|\tilde{X}| \sqrt{n_0 n_1},\quad M_{33}({\bf k}) = E_{+}({\bf k} ) - \mu,\quad M_{36} = - i \tilde g |\tilde{X}| n_0, \quad M_{55}({\bf k})=E_{-}({\bf k} ) - \mu.
    \end{align*}
\end{subequations} 

It is worth noting that $\hat{M}^{a(b)}$ is in general also momentum-dependent, since $\tilde{X}$ includes the excitonic Hopfield coefficients for the main and adjacent modes. However, we assume that in the limit of small $k$ when considering the first band only, 
one can approximately take $|\tilde X|\approx |{\bar X}_{0}^2 X_{+1}(-{\bf k}_0)X_{-1}({\bf k}_0)|$, 
so $\hat{M}^{a(b)}$ can be treated as constant.

\vor{To get insight into the low-energy fluctuation behavior, we truncate the matrix~\eqref{S12} to the $6\times 6$ matrix written in the fluctuations basis $\{u({\bf k}), v({\bf k}), u^{(+)}({\bf k} - {\bf k}_0), v^{(+)}({\bf k} - {\bf k}_0), u^{(-)}({\bf k} + {\bf k}_0), v^{(-)}({\bf k} + {\bf k}_0)\}$ considering $\omega \to 0$, $k \to 0$:
\begin{equation}\label{matrix1}
    \hat{M}^{\rm trunc}({\bf k})=
    \begin{pmatrix}
        M_{11}({\bf k}) & M_{12} &M_{13} & 0 & M_{13} & 0 \\ 
        -M^*_{12} & -M^*_{11}(-{\bf k})  & 0& M_{13} & 0 & M_{13}  \\
        M_{13}^* & 0 & M_{33}({\bf k}- {\bf k}_0) & 0 & 0 & M_{36} \\
        0 & M_{13}^* & 0 & -M^*_{33}(-{\bf k}- {\bf k}_0) & M_{36} & 0 \\
        M_{13}^* & 0 & 0 &  M_{36} & M_{55}({\bf k}+ {\bf k}_0) & 0 \\
        0& M_{13}^* & M_{36} & 0 & 0 &- M^*_{55} (-{\bf k}+ {\bf k}_0)
    \end{pmatrix}\!.
    \end{equation}}
\end{widetext}
\vor{In this limit, using the ansatz for the phase fluctuations
\begin{equation}
    \begin{pmatrix}
        u \\ v \\ u^{(+)} \\ v^{(+)} \\u^{(-)} \\ v^{(-)}
    \end{pmatrix} = \begin{pmatrix}
        \sqrt{n_0} i \delta \phi_0 \\ \sqrt{n_0}  i \delta \phi_0 \\ \sqrt{n_1} i \delta \phi_+ \\ \sqrt{n_1}  i \delta \phi_+ \\ \sqrt{n_1} i \delta \phi_- \\ \sqrt{n_1}  i \delta \phi_-
    \end{pmatrix}\!,
\end{equation}
we obtain the following equations for the phases $\delta \phi_0$, $\delta \phi_+$ and $\delta \phi_-$:
\begin{subequations}
    \begin{align}
    \frac{\hbar^2 k^2}{2 |m|}\delta \phi_0 + 2 i \tilde{g}|\tilde{X}|n_1 (2\delta \phi_0 - \delta \phi_+-\delta \phi_-) & = \hbar \omega \delta \phi_0, \nonumber \\
    i \tilde{g} |\tilde{X}| n_0(2\delta \phi_0 - \delta \phi_+-\delta \phi_-) \pm \hbar v k \delta\phi_{\pm} & = \hbar\omega \delta \phi_{\pm}, \nonumber
\end{align}
\end{subequations}
where approximations~\eqref{E_approx} and $m=-|m|$ were taken into account. 
Combining, one can straightforwardly obtain two coupled equations for $\delta \phi_0$ and $\delta \Phi = \delta \phi_+  -\delta \phi_-$:
\vspace{-10pt}
\begin{subequations}
    \begin{align}
        & \hbar \omega \delta \Phi = \hbar v k \biggl[ 2 -\frac{\hbar \omega + \frac{\hbar^2 k^2}{2 |m|}}{2i \tilde{g}|\tilde{X}| n_1}\biggr] \delta \phi_0,\\
        & \left[\! \hbar \omega  \biggl(2 - \frac{\hbar \omega + \frac{\hbar^2 k^2}{2 |m|}}{2i \tilde{g}|\tilde{X}|n_1} \biggr) \!-\! \frac{n_0}{n_1} \! \left( \! \hbar \omega + \frac{\hbar^2 k^2}{2 |m|}\right)\!\right] \! \delta \phi_0  = \hbar v k \delta \Phi
    \end{align}
\end{subequations}
The solution has the form:
\begin{multline}
    \hbar\omega \approx \pm i \hbar v k \sqrt{\frac{2 n_1}{n_0 - 2n_1}} - \frac{\hbar^2 k^2 }{2|m|}\frac{n_0}{n_0 - 4 n_1} \\
    - i \frac{\hbar^2 k^2 v^2}{2\tilde{g}\tilde{|X|}}\frac{n_0}{(n_0 - 4 n_1)(n_0-2n_1)} + \mathcal{O} (k^3),
\end{multline}
which in the lowest order of the low $k$ expansion yields Eq.~(9) of the main text.}\\

\vspace{10pt}
\section{Model applicability}

\subsection{\vor{The Bogoliubov approximation}}
\vspace{-5pt}
\vor{In the theory developed here, we consider the case of zero temperature and disregard the non-condensate corrections. Whereas in the analytical calculations the system is treated as one-dimensional and the finite size effects are not addressed, in numerical simulations presented in the main text, the characteristic scales of the condensate reproduce those reported in Ref.~\cite{trypogeorgos2025emerging}: the size of the condensate is approximately $50~\mu$m along $x$ and 
a few $\mu$m in transverse direction. 
With the parameters used in calculations ($n_1 = 0.1~\mu$m$^{-2}$, $n_0=520~\mu$m$^{-2}$), 
these condensate dimensions correspond to the 
total number of particles $N_{\rm tot} = \int d{\bf r}|\Psi_0 + \Psi_+ + \Psi_-|^2 \approx 10^6$ which is macroscopically large, ensuring validity of the mean-field approximation.
}

\subsection{\vor{Approximation for the reservoir density}}
\vspace{-5pt}
\vor{Having highlighted the role of reservoir interactions in renormalizing the interaction constant $g_{\rm eff}$ in the model of Eqs.~(2), it is of a great importance to confirm that when the adiabatic approximation for the reservoir density isn't approximated to first order but is kept in its full shape
\vspace{-10pt}
$$n_R = \frac{W}{R(1 + \eta |\Psi_0|^2)},$$
there is no qualitative change in the reported results. }

\vor{In this case, Eqs.~(2) of the main text become more intricate:}
\vspace{-5pt}
\begin{subequations}\label{GPE_new}
\begin{align}
     &\vor{i\hbar\frac{\partial}{\partial t}\Psi_{0}({\bf r}, t) = \varepsilon(\hat{\bf k})\Psi_{0}({\bf r}, t) + g|X_0|^4|\Psi_{0}|^2 \Psi_{0} } \nonumber \\
     &\qquad\quad \vor{+ \frac{g_R W \Psi_0}{R(1 \!+\! \eta |\Psi_0|^2)} \!+ 2 \tilde{g} \tilde{X} \Psi_{0}^* \Psi_{-}\Psi_{+} \!\!+ \frac{i W \Psi_{0}}{1 \!+\! \eta |\Psi_0|^2},} \label{GPE11} \\
     &\vor{i \hbar \frac{\partial}{\partial t}\Psi_{\pm}({\bf r}, t) =}  \vor{E_{\pm}(\hat{\bf k})\Psi_{\pm} + \tilde{g} \tilde{X}^{*} \Psi_{\mp}^*  \Psi_{0}^2,} \label{GPE21}
\end{align}
\end{subequations}
\vor{with no explicit change of the coefficient accompanying the interaction term $\sim|\Psi_0|^2\Psi_0$ in Eq.~\eqref{GPE11}. Furthermore, this equation does no longer have the Gross-Pitaevskii shape where each term could be given a transparent physical interpretation. Still, we performed the same analysis for the equations~\eqref{GPE_new} as given in Secs.~\ref{SMNote2-}--\ref{SMNote3-} here and  the End Matter (EM). The results presented in the main text do not suffer any qualitative change, except for the shifting values of the second threshold power $W_{\rm th2}$ and the stationary densities. The stability analysis of the Lyapunov exponents (Fig.~\ref{SFig2}) remains fully valid, and the long-wavelength analysis (see EM) is unaffected.}

\vor{Importantly, on the fluctuation level the stabilization mechanism with the growth of pumping is recovered. In particular, for the fluctuation matrices of the main diagonal $\hat{M}^{d}$ (see Sec.~\ref{SMNote3-}) we obtain:
\begin{subequations}\begin{align}
    & M_{11}({\bf k}) \to \varepsilon({\bf k}) - \mu + i \frac{W}{(1 + \eta n_0)} + g_R \frac{W}{R (1 + \eta n_0)} \nonumber  \\
    &\hspace{45pt}+ 2 g |X_0|^4 n_0 - g_R \frac{W n_0}{\gamma_R (1 \!+\! \eta n_0)^2} - 
    \frac{i\eta n_0 W}{(1 + \eta n_0)^2},\nonumber 
    \\
    & M_{12} \!\to  g |X_0|^4 n_0 \!-\! g_R \frac{W n_0}{\gamma_R (1 \!+\! \eta n_0)^2} \!-\! \frac{i \eta n_0 W}{(1 \!+\! \eta n_0)^2} \!+\! 2i \tilde{g}|\tilde X| n_1.\nonumber 
\end{align}
\end{subequations}
From both presented matrix elements, one sees that for $W > g|X_0|^4\gamma_R(1 + \eta n_0)^2/g_R$, interactions do become effectively attractive, which reproduces the stabilization condition for the model of the main text with the additional factor $(1 + \eta n_0)^2$. }

\vor{We conclude that, while retaining the full shape of $n_R$ will slightly alter the values of the fitting parameters that reproduce the experiment (such as $R$, $\gamma_R$, etc.), it does not change the reported physics. Yet, the gain-approximation model of Eqs.~(2), while being simplified, allows for a transparent interpretation of the stabilization mechanism.}

\subsection{\vor{Other model modifications}}

\vor{As discussed above, the model chosen in this work corresponds to the most intuitively transparent description that would, on the one hand, contain all the important ingredients of the system---such as the OPO mechanism, the three-mode order parameter which allows for the spontaneous breaking of the two $U(1)$ symmetries, and the interaction renormalization,---and, on the other hand, not overload the equations with additional nonlinearities. We note, however, that other contributions are possible, both in the Hamiltonian~(1) and the coupled GPEs~(2) of the main text. In particular, 
on the mean-field level, any additional condensate-induced or reservoir-induced blueshift of the adjacent $\pm1$ modes was neglected. Such modifications may result in (i) the change of the condensate energy $\mu$ and (ii) the redefined condition for the $\pm1$ modes momentum $\mp{\bf k}_0$, since the additional blueshift leads to the renormalization of the adjacent modes dispersion $E_{\pm}(\mp{\bf k}_0)\to E_{\pm}(\mp{\bf k}_0) + g |X_0|^2 |X_{\pm1}(\mp {\bf k}_0)|^2n_0$ or $E_{\pm}(\mp{\bf k}_0)\to E_{\pm}(\mp{\bf k}_0) + g_R^\prime n_R$, where $g_R^\prime\approx g |X_{\pm1}(\mp k_0)|^2$ is the $\pm1$-mode--reservoir interaction constant. Importantly, the threshold values~\eqref{th1}--\eqref{stationary} and the long-wavelength analytics do not change, and the fluctuation matrix does not qualitatively differ  from that presented in Sec.~\ref{SMNote3-}. As a consequence, the excitation spectra in such modified models do not substantially differ from those shown in Figs.~3 and~4(a) of the main text.}\\

\,\\